\documentclass[aps,prl,twocolumn,longbibliography,superscriptaddress,final,amsmath,amssymb]{revtex4-2}
\usepackage{graphicx}% Include figure files
\usepackage{dcolumn}% Align table columns on decimal point
\usepackage{bm}% bold math
\usepackage{hyperref}% add hypertext capabilities
\usepackage[usenames,dvipsnames]{xcolor}

\definecolor{vibrant}{HTML}{E77500}
\definecolor{muted}{HTML}{994400}

\hypersetup{
    colorlinks,
    linkcolor = vibrant,
    citecolor = muted,
    urlcolor = muted
}

\usepackage{braket}
\usepackage{enumitem}

\usepackage[english]{babel}

\newcommand{\tr}{\mathrm{Tr}}

\newcommand{\al}[1]{\begin{align}#1\end{align}}
\newcommand{\eq}[1]{\begin{equation}#1\end{equation}}

\newcommand{\Z}{\mathbb{Z}}
\newcommand{\R}{\mathbb{R}}

\newcommand{\I}{\mathbf{1}}

\renewcommand{\j}{\mathrm{i}}

\newcommand{\Span}{\mathrm{span}}

\newcommand{\re}{\mathrm{Re}}

\newcommand{\hc}{\mathrm{H.c.}}

\newcommand{\argdist}{\mathrm{argdist}}

\newcommand{\vect}[1]{\boldsymbol{#1}}

\usepackage[normalem]{ulem}

\newcommand{\OSC}{\rho^{\mathrm{osc}}}
\newcommand{\NESS}{\varrho^{\mathrm{\infty}}}
\newcommand{\NESSRDM}{\varrho^{\mathrm{\infty}}}
\newcommand{\sigmaRDM}{\sigma^{\mathrm{\infty}}}

\newcommand{\NESSpertRDM}{{\NESSRDM_\mathrm{pert}}}
\newcommand{\initRDM}{\varrho_\mathrm{init}}

\newcommand{\Liou}{\mathcal{L}}
\newcommand{\Lint}{\Liou_\mathrm{I}}

\begin{document}

\title{Quantum Synchronization of Perturbed Oscillating Coherences}

\author{Yi J. Zhao}
\affiliation{Department of Physics, University of California, Berkeley, California 94720, USA}

\author{Joel E. Moore}
\affiliation{Department of Physics, University of California, Berkeley, California 94720, USA}
\affiliation{Challenge Institute for Quantum Computation, University of California, Berkeley, California 94720, USA.}
\affiliation{Lawrence Berkeley National Laboratory, Berkeley, CA 94720, USA}

\author{Juzar Thingna}
\email{juzar18@gmail.com}
\affiliation{American Physical Society, 100 Motor Parkway, Hauppauge, New York 11788, USA}
\affiliation{Center for Theoretical Physics of Complex Systems, Institute for Basic Science (IBS), Daejeon 34126, Republic of Korea}

\author{Christopher W. W{\"a}chtler}
\email{cwwaechtler@icmm.csic.es}
\thanks{J.T. and C.W.W. contributed equally to this work.}
\affiliation{Instituto de Ciencia de Materiales de Madrid ICMM-CSIC, Madrid 28049, Spain}
\affiliation{Department of Physics, University of California, Berkeley, California 94720, USA}

\date{\today}

\begin{abstract}
Quantum mutual synchronization has recently been explored through the persistent oscillation of local observables that arises from undamped eigenmodes of dissipative dynamics. However, these oscillating modes require strictly fine-tuning the system to satisfy algebraic constraints. Here, we investigate the robustness of synchronization against generic perturbations that break these constraints. We identify conditions under which the steady state of the perturbed system exhibits correlations that indicate mutual synchronization, even as the oscillations decay. That synchronization persists as imprints in the time-independent, asymptotic steady state directly bridges the dynamical notion of synchronization with the steady-state notion, which have so far been treated as distinct phenomena. Moreover, we discover in a spin-1 model that the resulting steady-state synchronization is manifested in unexpected geometries of locked phases that are multiples of $\pi/3$. Our work establishes a link between the two primary paradigms of quantum synchronization while demonstrating its inherent robustness against generic perturbations.
\end{abstract}

\maketitle

\textit{Introduction.}---As one of the striking manifestations of collective behaviors, self-organized synchronization appears in diverse systems ranging from nearby placed pendulum clocks to cicadas with rhyming sounds~\cite{rosenblumSynchronizationPendulumClocks2003a, czeislerBrightLightResets1986, mooreClockAges1999}. The universality of synchronization in driven-dissipative systems makes it a central paradigm for studying complex dynamics and has inspired modern engineering~\cite{Pikovsky01, HerpichPRX18, strogatzNonlinearDynamicsChaos2024}. Synchronization in quantum systems, bearing potential application to quantum technologies, has moved to the forefront of theoretical~\cite{LS13,WNB14, WNB15, Loerch17,ishibashiOscillationCollapseCoupled2017, Sonar18, Bruder18_mutual, dutta2019critical,  thomasQuantumSynchronizationQuadratically2022, eshaqi-saniSynchronizationQuantumTrajectories2020, delmonteQuantumEffectsSynchronization2023a, shenEnhancingQuantumSynchronization2023, schmolkeMeasurementinducedQuantumSynchronization2023, xuSynchronizationTwoEnsembles2014a, Waechtler23, Tindall20,Buca22,Waechtler24, schmolkeNoiseinducedQuantumSynchronization2022, Sterba23, RVS23} and experimental~\cite{laskarObservationQuantumPhase2020c, koppenhoferQuantumSynchronizationIBM2020b, liExperimentalRealizationSynchronization2025b, taoNoiseinducedQuantumSynchronization2025, Liu25} investigations in the past decade.

Historically, theories of quantum synchronization have largely lain on two distinct conceptual grounds. Several seminal works identify synchronization within the time-independent steady state  of the dissipative dynamics~\cite{WNB15, Loerch17,ishibashiOscillationCollapseCoupled2017, Sonar18, Bruder18_mutual}. In this framework, a system exhibits \textit{steady-state synchronization} (SSync) if its stationary state shows correlated phases between subsystems~\footnote{Another paradigm dubbed \textit{entrainment}, where one of the subsystems may be viewed as an external signal of a known frequency, is also studied~\cite{WNB14}.}. For example, a pair of spins are considered synchronized if they exhibit a tendency to align while each spin individually has no preferred direction, i.e., its absolute phase remains free (as a quantum limit cycle). If an external field pins every spin to a definite direction, the system is trivially aligned but not synchronized.

In contrast, a more recent framework re-establishes the classical intuition of motion by defining synchronization through the real-time dynamics of local observables~\cite{Tindall20,Buca22,Waechtler24, schmolkeNoiseinducedQuantumSynchronization2022, Sterba23}. In this picture, a system exhibits \textit{dynamical synchronization} (DSync) if the expectation values of local observables, for generic initial conditions, oscillate persistently at an identical frequency and with a fixed phase difference. However, in the realm of open quantum systems governed by Markovian master equations, such sustained temporal oscillations arise from oscillating coherences, which are non-decaying modes of motion that exist only under strict algebraic constraints.~\cite{schmolkeNoiseinducedQuantumSynchronization2022, Buca22, Tindall20, Waechtler24}. Because these constraints also enforce multiple degenerate steady states, DSync appears structurally distinct from the earlier notion of SSync (in terms of correlations in the unique, time-independent steady state). Despite their disparate theoretical grounds, both DSync and SSync have been observed in recent quantum experiments~\cite{laskarObservationQuantumPhase2020c, koppenhoferQuantumSynchronizationIBM2020b, liExperimentalRealizationSynchronization2025b, taoNoiseinducedQuantumSynchronization2025, Liu25}. This apparent phenomenological overlap raises a fundamental question: Are these two primary notions of quantum synchronization related, or are they genuinely distinct physical phenomena?

\begin{figure}
    \centering
    \includegraphics[width=0.82\linewidth]{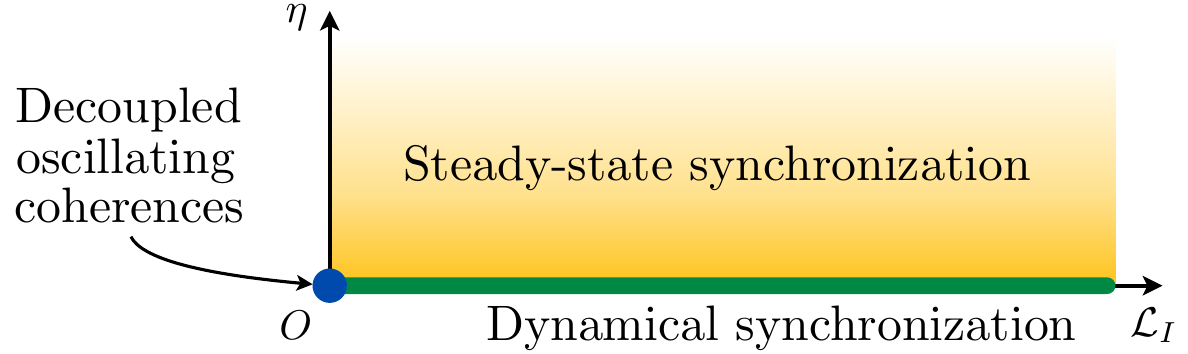}
    \caption{Dynamical synchronization (green line) typically features an engineered coupling $\Lint$ that fulfills specific algebraic constraints. The limit $\Lint \rightarrow 0$ leads to decoupled oscillating coherences (blue dot). Under generic perturbations of strength $\eta > 0$, the system possesses a single steady state exhibiting SSync (yellow region).} 
    \label{fig:regimes_cartoon}
\end{figure}

In this Letter, we bridge this gap by uncovering signatures of SSync within systems originally designed to exhibit DSync. We demonstrate that, while generic perturbations inevitably cause the oscillating coherences to decay (thereby excluding DSync), the system invariably relaxes into a unique final steady state that retains the phase correlations characteristic of SSync (see Fig.~\ref{fig:regimes_cartoon}). We analytically prove this connection for a general setup of weakly coupled oscillating coherences and verify it for established, strongly coupled models. Unexpectedly, in a perturbed spin-1 model, we find that the resulting SSync exhibits stable phase-angle differences in multiples of $\pi/3$---a rotational symmetry absent in the unperturbed model of DSync. Our results establish a link between the dynamical and steady-state paradigms, suggesting that synchronization is a robust phenomenon: It leaves an indelible static imprint even after its dynamic signature has decayed.

\textit{General setup.}---We focus on open quantum systems with a finite-dimensional Hilbert space $\mathbf{H}$. The dynamics of the reduced density matrix $\varrho$ is governed by the Markovian Lindblad master equation $\partial_t \varrho = \mathcal{L}[\varrho]$, where
\eq{
    \mathcal{L}[\varrho] = -\mathrm{i}\left[H, \varrho\right] + \sum_\mu \left(L_\mu \varrho L^\dagger_\mu - \frac{1}{2}\left\{L_\mu^\dagger L_\mu, \varrho\right\}\right). \label{equ:def_Liouvillian}
}
Here, $H$ denotes the system Hamiltonian and $L_\mu$ are the jump operators mediating dissipation. The Liouvillian $\mathcal{L}$ acts as a linear super-operator on the space of operators $\mathbf{B}=\left\{\ket{\psi}\bra{\phi}:\ket{\psi}, \ket{\phi}\in \mathbf{H} \right\}$. The eigenvalues $\lambda$ of $\mathcal{L}$ are generally complex, whose real parts lead to decay; $\re[\lambda] \leq 0$. We refer to the eigenmodes with $\lambda = 0$ as \textit{zero modes}. The physical steady states $\NESSRDM$ are constructed from linear combinations of these zero modes such that the resulting operator is a valid density matrix, being positive semidefinite (thus Hermitian) and having unit trace. Moreover, due to the structure of Eq.~\eqref{equ:def_Liouvillian}, if $\rho$ is an eigenmode with eigenvalue $\lambda$, then $\rho^\dagger$ is also an eigenmode with the complex conjugate eigenvalue $\lambda^*$~\cite{Lindblad76,Burgarth13}.

\textit{Two notions of synchronization.}---The physical manifestation of an oscillating coherence~\cite{Albert14, Thingna16, Manzano02012018, Thingna21} $\OSC$ (with purely imaginary eigenvalue $\lambda = \j \omega$) is persistent oscillations. For a generic initial state $\initRDM$, the density matrix evolution includes a non-decaying term proportional to $\OSC e^{\j \omega t} + \hc$. Consequently, any physical observable $O$ that overlaps with this mode (i.e., $\tr[O \OSC]\neq 0$) exhibits persistent oscillations in its expectation value. If two such observables, $O_j$ and $O_k$ localized on disjoint subsystems $j$ and $k$, both overlap with the mode, their expectation values $\braket{O_j}(t)$ and $\braket{O_k}(t)$ will oscillate at the same frequency $\omega$, maintaining a locked phase difference that is independent of the initial state. This phenomenon is known as \textit{dynamical synchronization}~\footnote{In Ref.~\cite{SJHV22}, synchronization resulting from multiple pairs of oscillating coherences is referred to ``coherence synchronization,'' whereas in this work, in the spirit of Ref.~\cite{Buca22}, we consider the simple case of local observables being synchronized in the presence of a single pair of oscillating coherences.}.

In this work, we focus on $N$-spin systems where DSync is probed by a continuous family of locally rotated observables $O^\phi = e^{\j \phi S^z} O e^{-\j \phi S^z}$, parameterized by an azimuthal phase angle $\phi\in\R$. Here, $O$ represents an operator in the transverse $x$-$y$ plane [e.g., $S^x$ or $(S^x)^2$] chosen such that $\tr[O^\phi\OSC] \propto e^{\j n(\phi-\Phi)}$ for some nonzero integer $n$ and a reference phase $\Phi\in\R$~\footnote{We note that in other systems~\cite{waechtler25},  different forms of observables may arise, leading to a modified picture.}. To satisfy the \textit{limit-cycle requirement} that the absolute phase remains free, we strictly require that the expectation value of the observable evaluated on any zero mode, $\tr[O^\phi \rho^{\lambda=0}_\alpha]$, be independent of $\phi$. This setup can be found in the dynamical synchronization models reported in Refs.~\cite{Tindall20, Buca22}. Physically, these conditions ensure that the observables evolve as $\braket{O^{\phi_j}_j}(t) = A_j \cos[\omega (t-t_0) + n(\phi_j - \Phi_j)] + B_j$. While the amplitudes $A_j$, baseline shifts $B_j$, and temporal offset $t_0$ depend on the initial state, the crucial phase offsets $\Phi_j$ do not. In this picture, DSync indicates that the free azimuthal spin angles $\phi_j$, $\phi_k$ rotate collectively at frequency $\omega/n$, maintaining phase differences $\Phi_j - \Phi_k$.

In stark contrast, \textit{steady-state synchronization} does not rely on persistent temporal dynamics. It is formally defined in systems possessing a unique asymptotic state $\NESSRDM$ (such that there is only one zero mode, into which every initial state decays). Here, synchronization is identified via static spatial correlations in $\NESSRDM$: The subsystems exhibit a statistical propensity to align at a fixed relative phase. To connect SSync with DSync, we probe these static correlations with the same operator family $O^\phi$ utilized in the dynamical picture. A state exhibits SSync if a multi-point correlation function $\tr[O^{\phi_{j_1}}_{j_1} \cdots O^{\phi_{j_m}}_{j_m} \NESSRDM]$ exhibits maxima at some configurations $(\phi_{j_1}, \cdots, \phi_{j_m})$ of azimuthal angles, thereby favoring specific relative phase differences. Crucially, the angles must remain locally free; the reduced local state of any single spin must exhibit no preference for $\phi$~\footnote{For two-level systems, there remains active debate regarding the precise definition, and even the existence of, true limit cycles~\cite{Bruder18_driven, parra-lopezSynchronizationTwolevelQuantum2020a, Kurzynski20, Zhang23}. However, within the context of SSync, a phase-invariant marginal distribution is widely accepted as the quantum analog of a free classical phase variable.}. Technically, these phase preferences are quantified by the joint probability distribution $S_\mathrm{d}(\boldsymbol{\phi}')$ of the phase differences $\boldsymbol{\phi}' = (\phi_1 - \phi_N, \cdots \phi_{N-1} - \phi_N)$, obtained by marginalizing the Husimi-Q phase-space representation~\cite{Bruder18_driven, Bruder18_mutual,Thingna23, Murtadho23}. Isolated maxima in $S_\mathrm{d}$ indicate SSync, whereas a diagonal state in the computation basis yields a uniform $S_\mathrm{d}$ indicating no synchronization~(see Appendix~\ref{appendix:A}).

\textit{Decoupled oscillating coherences.}---As a foundational step in bridging DSync and SSync, we first consider a  non-interacting system of $N$ identical, decoupled spins, each independently hosting a pair of oscillating coherences (represented by the blue dot in Fig.~\ref{fig:regimes_cartoon}). We will analytically demonstrate that the degenerate steady-state manifold of this non-interacting system necessarily contains at least one steady state that exhibits correlations in the same operator that probes the oscillating coherences. Later, we will show that when generic perturbations are introduced to couple the spins, the system collapses into a unique steady state that locks into these correlations.

We set up every single spin as follows. Let the Liouvillian be $\mathcal{L}$ and possess a pair of oscillating coherences $\OSC$, $\left(\OSC\right)^\dagger$ with eigenvalues $\pm \j \omega$, respectively, and steady states $\NESS_{\alpha}$, with no other zero modes. Let $O^\phi$ denote the locally rotated single-spin observables; they overlap with the oscillating coherence, such that $\tr[O^\phi \OSC] = A e^{\j n\phi}$ for some real number $A > 0$ and integer $n$. On the other hand, we impose the limit-cycle requirement, namely that $\tr[O^{\phi} \NESS_{\alpha}]$ be independent of the azimuthal phase $\phi$. For future reference, we identify two special linear combinations $\sigma_1^\infty$ and $\sigma_2^\infty$ of the steady states $\NESS_{\alpha}$, such that $\sigma_1^\infty = U  \OSC$ and $\sigma_2^\infty = \OSC U$ for some unitary matrix $U$ (which exists as we show in the Supplemental Material (SM)~\cite{SM}, based on Ref.~\cite{Buca22}). Define $B_\beta = \tr[O^{\phi} \sigmaRDM_{\beta}]$, for $\beta = 1, 2$, which are then also $\phi$-independent. For brevity, we will focus on $N=2$ identical copies of such a spin; similar analysis carries over to $N > 2$.

The critical insight lies in the structure of the joint zero modes of the whole decoupled system: They include not only the steady states $\NESS_{\alpha, \beta} = \NESS_{\alpha, \mathrm{A}} \otimes \NESS_{\beta, \mathrm{B}}$ but also nontrivial cross-terms formed by the oscillating coherences: $\rho^{\lambda=0}_\mathrm{new} = \OSC_{\mathrm{A}} \otimes \left(\OSC_{\mathrm{B}}\right)^\dagger$ and $(\rho^{\lambda=0}_\mathrm{new})^\dagger$. Here, the subscripts $j = \mathrm{A}, \mathrm{B}$ label the spins. One can verify that $(\mathcal{L}_\mathrm{A}\otimes \mathcal{I}_\mathrm{B} + \mathcal{I}_\mathrm{A} \otimes \mathcal{L}_\mathrm{B})[\rho^{\lambda=0}_\mathrm{new}] = 0$, where $\mathcal{I}: \rho\mapsto\rho$ is the identity channel on a single spin. We can then construct a family of well-defined~\cite{SM}, globally correlated steady states parameterized by a complex amplitude $z = u e^{i\xi}$ with $0< u < 1$ and an arbitrary phase $\xi\in\R$,
\eq{
    \NESSRDM_z = \frac{1}{2}\left[\sigmaRDM_{1, \mathrm{A}} \otimes \sigmaRDM_{2, \mathrm{B}} + \sigmaRDM_{2, \mathrm{A}} \otimes \sigmaRDM_{1, \mathrm{B}} + \left(z\rho^{\lambda=0}_\mathrm{new} + \hc\right)\right]. \label{equ:mixture_twosites}
}
Crucially, the two-point correlation function reads $\tr[(O_\mathrm{A}^{\phi_\mathrm{A}} \otimes O_\mathrm{B}^{\phi_\mathrm{B}}) \NESSRDM_z] = uA \cos[n(\phi_\mathrm{A}-\phi_\mathrm{B}) + \xi] + B_1 B_2$. The cosine term indicates that, within the state $\NESSRDM_z$, the local azimuthal angles are correlated, statistically favoring phase differences of $\phi_\mathrm{A} - \phi_\mathrm{B} = (2m\pi - \xi)/n$ ($m\in\Z$). Concurrently, the reduced state of each individual spin remains uniform; for example, $\tr[(O_\mathrm{A}^{\phi_\mathrm{A}}\otimes \I_\mathrm{B})\NESSRDM_z]$ evaluates to a $\phi_\mathrm{A}$-independent constant. Therefore, the state $\NESSRDM_z$ by itself features SSync. However, in this unperturbed, degenerate scenario, the specific steady state $\NESSRDM_z$ (and thus the resultant phase difference) that the system relaxes into depends entirely on the initial state.

\textit{Perturbing oscillating coherences.}---In any realistic experimental architecture, isolated oscillating coherences are an idealization; systems inevitably suffer from weak internal interactions or environmental decoherence. Generic perturbations of these sorts explicitly break the delicate algebraic constraints required to sustain purely imaginary eigenvalues. Consequently, the oscillating coherences acquire a nonzero real part and decay exponentially at long times, which destroys DSync. Simultaneously, the generic perturbations split the degeneracy of the zero-mode manifold, forcing the system to relax into a unique steady state $\NESSpertRDM$~\cite{Evans77}.

Under degenerate perturbation theory, the unique perturbed steady state can be expanded to leading order as a linear combination of the unperturbed zero modes:
\eq{
    \NESSpertRDM = \sum_\alpha c_\alpha  \rho^{\lambda=0}_\alpha. \label{equ:def_mixture_NESS}
}
The specific coefficients $c_\alpha$ are dictated by the explicit form of the perturbation and constrained only by the requirement that $\NESSpertRDM$ remains a valid density matrix.

Remarkably, Eq.~\eqref{equ:def_mixture_NESS} indicates that the system will exhibit robust SSync \textit{after} the perturbation (i.e., after the dynamical oscillations have decayed), provided two conditions are met in the unperturbed system. (i) The local azimuthal angle of every unperturbed zero mode $\rho^{\lambda=0}_\alpha$ must be free. If true, the unique state $\NESSpertRDM$ will inherit this lack of local preference, satisfying the \emph{limit-cycle} requirement. (ii) At least one of the unperturbed steady states must exhibit nontrivial phase correlations. Unless the perturbation is maliciously fine-tuned to exactly cancel specific coefficients in Eq.~\eqref{equ:def_mixture_NESS}, these correlated states will generically contribute to the expansion of $\NESSpertRDM$.

As we have shown with Eq.~\eqref{equ:mixture_twosites}, the decoupled system satisfies both conditions. Therefore, when generic perturbations are applied (moving vertically away from the blue dot in Fig.~\ref{fig:regimes_cartoon}), the unique final steady state $\NESSpertRDM$ will contain a nonzero contribution from the correlated cross-terms. The specific perturbation selects a definite phase offset $\xi$, thereby locking the spins into an initial-state-independent relative phase difference. We detail a two-spin model showing this mechanism in the SM~\cite{SM}.

This mechanism for SSync extends naturally to strongly coupled models of DSync. These models rely on engineered dissipative couplings $\Lint$ (the green line in Fig.~\ref{fig:regimes_cartoon}) to sustain collective oscillations. By smoothly tuning $\Lint \rightarrow 0$, the system continuously connects to the decoupled limit. The correlations of Eq.~\eqref{equ:mixture_twosites} do not instantaneously vanish when the coupling $\Lint$ is finite. Consequently, when generic perturbations break the idealized $\Lint$ constraints, the system again collapses into a unique steady state that inherits these remnant correlations, manifesting SSync (the yellow region in Fig.~\ref{fig:regimes_cartoon}). We now illustrate this behavior in an interacting model.

\textit{Spin-$1$ model.}---Consider a paradigmatic, strongly coupled model of DSync~\cite{Tindall20}, consisting of $N$ spin-1s. The coherent dynamics is governed by the Hamiltonian
\eq{
    H = \omega \sum_{j=1}^N S_j^z + \sum_{j=1}^{N-1} \frac{J}{2}\left(S_j^+ S_{j+1}^- + \hc\right) + \Delta S_j^z S_{j+1}^z, \label{equ:spin1:H}
}
while dissipation is driven identically on each site by the local jump operators $ L_j = \sqrt{\gamma} \left(S_j^z\right)^2$. Here, $S^\pm = S^x \pm \j S^y$ are the standard spin ladder operators, and $S^z_j\ket{\mathbf{m}} = m_j\ket{\mathbf{m}}$ defines the computation basis states, where $\mathbf{m} = (m_1, \cdots, m_N)$. Note that the decoupled limit ($\Lint \rightarrow 0$) corresponds to smoothly taking the interaction parameters $J, \Delta \rightarrow 0$.

We study the model analytically, noting a strong~\cite{BP12} $\mathrm{U}(1)$ symmetry generated by $M = \sum_j S_j^z$ with eigenvalues $-N,\cdots, N$. A strong symmetry divides the Hilbert space such that a state evolves within its initial sector (see Appendix~\ref{appendix:B}).  The $M=0$ sector is further divided into two by a strong $\Z_2$ symmetry of global spin inversion, $P = \sum_\mathbf{m} \ket{-\mathbf{m}}\bra{\mathbf{m}}$. We denote the two sectors by $0\pm$, characterized by the eigenvalues $P=\pm 1$, respectively. Note that $PM=-MP$, so the overall symmetry is the semi-direct product $\mathrm{U}(1)\rtimes\Z_2$. We expect $2(N+1)$ linearly independent steady states $\NESSRDM_\alpha$, which we label with the quantum numbers $\alpha = 0\pm, \pm 1,\cdots, \pm N$. Furthermore, as all the jump operators $L_j$ are Hermitian, the steady state within each sector is the maximally mixed state: $\NESSRDM_\alpha \propto \I_\alpha$, where $\I_\alpha$ is the projector onto sector $\alpha$ (see Appendix~\ref{appendix:B}). Crucially, the projectors $\I_M$ are diagonal in the computation basis for $M \neq 0$. Only $\I_{0\pm}$ contain off-diagonal coherences.

On the other hand, the persistent oscillations defining DSync arise from the coherent resonances between the $\pm M$ sectors (for $M\neq 0$)~\cite{Tindall20}. In particular, the oscillating coherences are the eigenmodes $\OSC_M = P\NESS_M = P \I_M$ with purely imaginary eigenvalues $2\j\omega M$; one can verify this directly, noting that the relation $PM=-MP$ leads to $[P, H] = 2\omega PM$. To observe the oscillations, a local operator $O_j$ must hybridize the $\pm M$ sectors of inverse total magnetizations. The squared transverse spin operator $O_j = (S^x_j)^{2}$ fulfills this role; it overlaps with the oscillating coherence at $M=1$, yielding $\tr[O_{j}^\phi\OSC_1] \propto e^{2\j \phi}$. Additionally, $\tr[O_{j}^\phi \rho^\infty_\alpha]$ for every $\rho^\infty_\alpha \propto \I_\alpha$ picks up only the diagonal entries of $O_{j}^\phi$~(cf. Appendix~\ref{appendix:B}), which are independent of $\phi$. The $\mathrm{U}(1)\rtimes\Z_2$ symmetry therefore promises DSync, which is probed by the operators $O_{j}$.

We now analytically show that, when generic perturbations break these delicate symmetries, the resulting unique steady state exhibits SSync. This requires verifying the two conditions that we have established with Eq.~\eqref{equ:def_mixture_NESS}. Condition (i) is satisfied because the local reduced density matrix for any single spin, derived by tracing out the rest of the spins from $\NESS_\alpha \propto \I_\alpha$, is strictly diagonal. Thus, the individual spin angles are free, satisfying the limit-cycle requirement. Condition (ii) is satisfied because each of the two steady states $\NESS_{0\pm} \propto \I_{0\pm}$ explicitly contains off-diagonal coherence terms. These off-diagonal terms yield a correlated two-point function: $\tr[O_{j}^{\phi_j} O_{k}^{\phi_k}\NESS_{0\pm}] \propto \cos[2(\phi_j - \phi_k)]$. Because these correlated $0\pm$ states generically contribute to the unique perturbed steady state expansion via Eq.~\eqref{equ:def_mixture_NESS}, the system will lock into these phase differences after all the oscillations have decayed. In SM~\cite{SM}, we generalize this model to spin-$s$ and argue that, for $s>1$ being integer (half integer), both DSync and SSync appear (do not occur).

We emphasize that this translation from DSync to SSync is rooted in the symmetries of the model; it is not a universally guaranteed consequence of generic Liouvillian structures. To demonstrate this, consider a modified variant of the spin-1 model containing an additional nonlocal jump operator $L' = \sqrt{\gamma'} S_1^x\prod_{\nu=1}^{N} (\nu-M)(\nu+M)$. By construction, $L'$ annihilates any state where $M \neq 0$. Therefore, it leaves the $M\neq 0$ sectors completely untouched, ensuring the oscillating coherences $\OSC_M$ remain perfectly intact. The model thus still exhibits DSync. However, $L'$ actively drives the probability amplitude out of the $M=0$ sectors,  destroying the correlated steady states $\NESS_{0\pm}$. Consequently, the remaining steady states $\NESSRDM_M\propto \I_M$ ($M\neq 0$) are all diagonal in the computation basis, lacking the static phase correlations required by condition (ii). In this pathological scenario, applying symmetry-breaking perturbations will destroy the DSync but will \textit{not} result in SSync. However, we note that this circumstance happens because the coupling $L'$ selects global sectors, which is impossible in models like the one given around Eq.~\eqref{equ:spin1:H} that contain only local terms.

\textit{The phase profile and $\pi/3$ SSync.---}Focusing on the $N=3$ case, we now demonstrate that the SSync of the spin-1 model can also be characterized by the widely used probability distributions of azimuthal spin angles (explicitly defined in Appendix~\ref{appendix:A}). According to Eq.~\eqref{equ:def_mixture_NESS} the steady state of concern, under generic perturbations, takes the form $\NESSpertRDM = \sum_\alpha c_\alpha \I_\alpha$ with non-negative, perturbation-dependent coefficients $c_\alpha$. Because only $\NESSRDM_{0\pm} \propto \I_{0\pm}$ contain off-diagonal elements [cf. Eq.~\eqref{equ:spin1:off-diagonals} in Appendix~\ref{appendix:B}], the joint distribution $S(\boldsymbol{\phi})$ of the azimuthal angles $\phi_1, \phi_2, \phi_3$ simplifies to
\eq{
    S(\bm{\phi}) = \frac{1}{(2\pi)^3} + \frac{c_{0+}-c_{0-}}{32\pi^3}\sum_{j=1}^3 \cos[2(\phi_j - \phi_{j+1})], \label{equ:S_phi_spin1}
}
where we denote $j+3\equiv j$. Because $S(\bm{\phi})$ is invariant under global phase shifts, we obtain the distribution of relative phase differences by fixing the third spin as a reference: $S_\mathrm{d}(\phi_1', \phi_2') = 2\pi S(\phi_1', \phi_2', 0)$, where $\phi_1' = \phi_1 - \phi_3$ and $\phi_2' = \phi_2 - \phi_3$.

\begin{figure}
    \centering
    \includegraphics[width=0.9\linewidth]{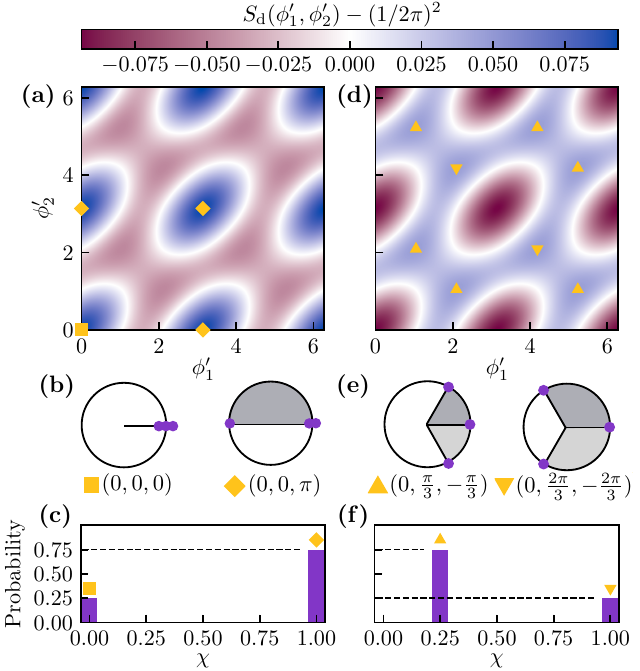}
    \caption{Analytically predicted SSync in the $N = 3$ spin-1 model. Panels {(a--c)} represent $(0, \pi)$ alignment while panels {(d--f)} show $\pi/3$ alignment. {(a)} and {(d)}: The two possible distributions $S_\mathrm{d}(\phi_1', \phi_2')$ [see below Eq.~\eqref{equ:S_phi_spin1}] of the angular differences $\phi_1' = \phi_1-\phi_3$, $\phi_2' = \phi_2-\phi_3$ for a single random sample. {(b)} and {(e)}: The configurations of the spin angles that maximize $S_\mathrm{d}$, with the corresponding maxima marked in {(a)} and {(d)} with yellow symbols. {(c)} and {(f)}: The $\chi$ values [Eq.~\eqref{equ:chi_phase_diff}] of the maxima and their occurrance rates in $S_\mathrm{d}(\phi_1', \phi_2')$.
    }
    \label{fig:S_phis}
\end{figure}

Equation~\eqref{equ:S_phi_spin1} reveals that the phase configuration underlying SSync is sensitive to the sign of $(c_{0+} - c_{0-})$. To analyze this, we refer to angular configurations as unordered tuples $(\phi_1, \phi_2, \phi_3) \equiv (\phi_1', \phi_2', 0)$ up to global phase shifts and permutations [e.g., $(0, 0, \pi)$ is equivalent to $(0, \pi, \pi)$]. If the perturbation yields $c_{0+} > c_{0-}$, then $S_\mathrm{d}$ is maximized when every cosine function in Eq.~\eqref{equ:S_phi_spin1} is maximized, i.e., at $(0, 0, 0)$ and $(0, 0, \pi)$, represented by the yellow symbols in Fig.~\ref{fig:S_phis}\textcolor{vibrant}{a} and on a circle in Fig.~\ref{fig:S_phis}\textcolor{vibrant}{b}. These maxima indicate that the spins prefer to lock in parallel or anti-parallel orientations,  which is in agreement with the DSync of $(S^x_j)^2$~\footnote{In particular, for spins whose azimuthal angles (i.e., the spherical angles in the plane perpendicular to the $z$-direction) differ by $\pi$ or $0$, we expect them to have $x$-components that either differ by a sign or agree. This picture agrees with the equivalent dynamics of the expectation value of $\left(S^x\right)^2$ in the dynamical notion.}.

However, a strikingly different scenario occurs if the perturbation yields $c_{0-} < c_{0+}$. In this regime, the system attempts to minimize the three coupled cosine functions in Eq.~\eqref{equ:S_phi_spin1}, which leads to geometric frustration. The resulting optimal phase-locked configurations are $(0, \pi/3, -\pi/3)$ and $(0, 2\pi/3, -2\pi/3)$, visualized in Figs.~\ref{fig:S_phis}\textcolor{vibrant}{d} and \ref{fig:S_phis}\textcolor{vibrant}{e}. Here, the spins synchronize with stable relative phase differences that are strict multiples of $\pi/3$. Such \textit{$\pi/3$-synchronization} is a novel feature of the static steady state, bearing no obvious physical signature in the unperturbed model of DSync. This demonstrates that shifting to the steady-state perspective can reveal significantly richer many-body correlations that are invisible to the DSync analysis~\footnote{For completeness, in the Supplemental Material~\cite{SM}, we analyze a separate interacting spin-1/2 model~\cite{Buca22}. In that system, the symmetry constraints ensure the steady-state phase differences exclusively lock at $\pi$, perfectly mirroring its dynamical behavior free from frustration effects.}.

\textit{Numerical verification.}---We sample random perturbations of strengths $\eta$ from an ensemble of random Liouvillians~\cite{SM}. For every sample, exact diagonalization yields a single steady state, from which we compute the distribution $S_\mathrm{d}(\phi_1', \phi_2')$ of the relative phase angles. To systematically classify the phase-locked configurations, we define the geometric characterization function
\eq{
\chi(\phi_1, \phi_2, \phi_3) = \min_{j \in \left\{0, 1, 2\right\}} \left[\frac{3}{4\pi}\sum_i \mathrm{argdist}\left(\phi_i, \overline{\phi}_j\right)\right]^2.
\label{equ:chi_phase_diff}
}
Here, $\overline{\phi}_j = \left(2\pi j + \sum_k \phi_k\right)/3$ represents the mean orientation, shiftable by $2\pi/3$ increments, and the function $\mathrm{argdist}\left(\theta, \psi\right) = \min_{n\in\mathbb{Z}} \left|\theta - \psi + 2\pi n\right|$ computes the shortest periodic distance between angles on the unit circle. The function $\chi$ is invariant under global rotations, effectively collapsing to $\chi(\phi_1', \phi_2', 0)$. Crucially, $\chi$ approaches $0$ for perfectly aligned angles, while spreading uniformly between $[0, 1]$ for uncorrelated, uniform angles~\cite{SM}. To isolate the synchronization features, we obtain the distribution of $\chi$ for values of $S_\mathrm{d}$ above $r \max S_\mathrm{d}$. We pick $r = 0.95<1$ to avoid potential numerical artifacts. See SM~\cite{SM} for detailed methods.

\begin{figure}
    \centering
    \includegraphics[width=0.99\linewidth]{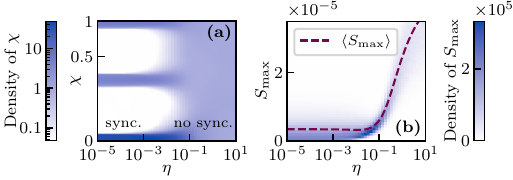}
    \caption{Numerical verification of SSync under random perturbation. (a) Average distribution of $\chi$ as sampled from $S_\mathrm{d}(\phi_1', \phi_2')$ above $0.95\max S_\mathrm{d}$, at varying perturbation strengths $\eta$. The vertical axis is on a scale such that $\sqrt{\chi}$ is uniformly spaced.  (b) The distribution of  $S_\mathrm{max} = \max S_\mathrm{d}(\phi_1', \phi_2')- (1/2\pi)^2$. We randomize over $10^4$ samples; $\omega = J = 2\Delta = \gamma/2 = 1$.}
    \label{fig:varying_eta}
\end{figure}

The ensemble-averaged results are presented in Fig.~\ref{fig:varying_eta}\textcolor{vibrant}{a}. To provide a theoretical baseline, we plot the exact analytical limits ($\eta \rightarrow 0$, $r\rightarrow 1$) for the standard $(0, \pi)$ alignment and the frustrated $\pi/3$ alignment in Figs.~\ref{fig:S_phis}\textcolor{vibrant}{c} and \ref{fig:S_phis}\textcolor{vibrant}{f}, respectively. The numerical data in Fig.~\ref{fig:varying_eta}\textcolor{vibrant}{a} demonstrates that for weak generic perturbations ($\eta / \omega \lesssim 10^{-3}$), the system locks into three distinct, sharp peaks. These peaks perfectly map onto the analytical signatures of both the (0,$\pi$)- and $\pi/3$-synchronization, proving that both types of synchronization naturally occur under random perturbations. Quantitative analysis~\cite{SM} confirms both occur at roughly equal statistical rates. As the perturbation strength $\eta$ becomes dominant, the phase-locking peaks smear out as expected, indicating the complete destruction of SSync.

Finally, analyzing in Fig.~\ref{fig:varying_eta}\textcolor{vibrant}{b} the distribution of the maximum synchronization magnitude $S_\mathrm{max} = \max S_\mathrm{d} - 1/4\pi^2$ calibrated by the uniform distribution, we find an abundance of nonzero values, which confirms that SSync persists in the weak-perturbation limit ($\eta / \omega \lesssim 10^{-3}$). While $S_\mathrm{max}$ technically increases at very high $\eta$, this is an artifact of randomly frozen angles, not an indication of enhanced synchronization (as evidenced by the complete absence of definite phase-locking patterns in panel \textcolor{vibrant}{a}). Our results indicate robust SSync in a finite vicinity ($0 < \eta/\omega \lesssim 10^{-3}$) of DSync.

\textit{Conclusion.}---In this Letter, we bridge the dynamical and steady-state frameworks of quantum synchronization. We demonstrate that the algebraic constraints required to sustain dynamic oscillations can give rise to static phase correlations. Consequently, even when generic perturbations cause these oscillations to decay, the system reliably relaxes into a unique, phase-locked steady state. Exploring this static perspective uncovers geometrically frustrated synchronization that is invisible in the unperturbed limit.  Our analysis serves as a foundational step toward a unified theory of quantum synchronization, highlighting that combining both perspectives yields significantly richer physics than either approach can capture alone.

\textit{Acknowledgments.}---The authors thank B. Bu{\v c}a for helpful comments. Y.J.Z. and J.E.M. acknowledge support from the National Science Foundation, Challenge Institute for Quantum Computation at UC Berkeley. C.W.W. was supported by the Deutsche Forschungsgemeinschaft (DFG, German Research Foundation), Project No. 496502542 (WA 5170/1-1) and received funding from the European Union's Horizon Europe research and innovation programme under the Marie Skłodowska-Curie Actions (MSCA) grant agreement No. 101149948. This research used the Lawrencium computational cluster resource provided by the IT Division at the Lawrence Berkeley National Laboratory (supported by the Director, Office of Science, Office of Basic Energy Sciences, of the U.S. Department of Energy under Contract No. DE-AC02-05CH11231). Views and opinions expressed are, however, those of the author(s) only and do not necessarily reflect those of the European Union or the European Research Council. Neither the European Union nor the granting authority can be held responsible for them.

\section*{END MATTER}

\newcounter{endmattersection}
\setcounter{endmattersection}{0}
\renewcommand{\theendmattersection}{\Alph{endmattersection}}
\newcommand{\heading}[1]{\refstepcounter{endmattersection}\textit{Appendix \theendmattersection: #1}}

\heading{Quantifying SSync.}\label{appendix:A}---We explicitly define the distribution $S(\boldsymbol{\phi})$ of azimuthal angles $\boldsymbol{\phi} = (\phi_1, \cdots, \phi_N)$ and the quantities derived thereof, including the distribution $S_\mathrm{d}$ of angular differences and a steady-state synchronization measure $S_\mathrm{max}$. Here, as is throughout the work, $N$ denotes the number of spins, which have quantum number $s$. We then show that $S_\mathrm{d}$ is uniform and thus $S_\mathrm{max}=0$ if the steady state $\NESSRDM$ is diagonal in the computation basis, in agreement with the intuition that diagonal states do not show synchronization. We remark that, rigorously, the distributions of interest are joint quasi-probability distributions because of the quantum nature of the steady state $\NESSRDM$, but we still call them \textit{distributions} for brevity.

The distribution $S(\boldsymbol{\phi})$ can be derived from the distribution $Q(\vect{\phi}, \vect{\theta})$ of spin spherical angles $(\vect{\phi}, \vect{\theta}) = (\phi_1,\cdots, \phi_N, \theta_1,\cdots, \theta_N)$. The latter is known as the Husimi-Q representation of the phase space~\cite{Husimi40},
\eq{
Q(\vect{\phi}, \vect{\theta}) = \left(\frac{2s+1}{4\pi}\right)^N\bra{\vect{\phi}, \vect{\theta}}\NESSRDM\ket{\vect{\phi}, \vect{\theta}}, \label{equ:Husimi_Q}
}
defined with generalized $\mathrm{SU}(2)$ coherent states~\cite{Nemoto00}
\eq{
\ket{\vect{\phi}, \vect{\theta}} = \bigotimes_j \left(e^{-\j \phi_j S^z_j} e^{-\j \theta_j S^y_j} \ket{s}_j\right).
}
Here, for every spin $j$, we refer to the local spin operators by $S_j^{z}$, $S_j^{y}$ and to the local state with the maximum positive magnetization in the $z$ direction by $\ket{s}_j$. One can check that Eq.~\eqref{equ:Husimi_Q} has the correct normalization as the maximally mixed state gives a uniform $Q(\vect{\phi}, \vect{\theta}) = 1/(4\pi)^N$. As long as the synchronization of azimuthal angles is concerned, the population angles $\vect{\theta}$ are irrelevant and thus integrated out. We then arrive at the distribution of the angles $\phi_j$,
\eq{
S(\vect{\phi}) = \prod_j\left(\int_0^\pi \sin \theta_j \, d\theta_j\right) Q(\vect{\phi}, \vect{\theta}). \label{equ:phase_distribution_def}
}
For free angles that each does not show a preferred direction, it suffices to consider the joint distribution of their phase differences $\phi_j' = \phi_j-\phi_N$ ($\forall j = 1,\cdots, N-1$)~\cite{Bruder18_driven, Bruder18_mutual}
\eq{
    S_\mathrm{d}(\boldsymbol{\phi}') = \int_0^{2\pi} d\phi_N\,S(\phi_1'+\phi_N,\cdots, \phi_{N-1}' + \phi_N, \phi_N). \label{equ:phase_diff_distribution_def}
}
Its maximum calibrated by the uniform distribution,
\eq{
S_\mathrm{max} = \max_{\boldsymbol{\phi}'} S_\mathrm{d}(\vect{\phi}') - \left(1/2\pi\right)^{N-1}, \label{equ:phase_measure_def}
}
has been taken as a measure for SSync~\cite{Bruder18_driven, Bruder18_mutual}.

We now show that, when the steady state $\NESSRDM$ is diagonal in the computation basis, the distribution $S(\boldsymbol{\phi})$ of azimuthal spin angles is uniform. Indeed, every diagonal entry contributes to $\NESSRDM$ a component proportional to $\rho_{\mathbf{m}, \mathbf{m}} = \ket{m_1 \cdots m_N}\bra{m_1 \cdots m_N}$, where $\mathbf{m} = (m_1, \cdots, m_N)$ and $m_j$ denotes the magnetization of the spin $j$, along the $z$ direction. Note that $e^{\j \phi_k S_k^z} \rho_{\mathbf{m}, \mathbf{m}} e^{-\j \phi_k S_k^z}$ does not have any $\phi_k$ dependence, and this holds for all $k$. Consequently, $Q(\vect{\phi}, \vect{\theta})$ and thus $S(\boldsymbol{\phi})$ do not depend on $\boldsymbol{\phi}$. Since $S(\boldsymbol{\phi})$ as a distribution normalizes to unity, namely $\left(\prod_j \int_{0}^{2\pi} d \phi_j\right) S(\boldsymbol{\phi}) = 1$, we have $S(\boldsymbol{\phi}) = (1/2\pi)^{N}$. The derived distribution $S_\mathrm{d}(\boldsymbol{\phi}')$ of angular differences [see Eq.~\eqref{equ:phase_diff_distribution_def}] is thus uniformly $(1/2\pi)^{N-1}$. It therefore follows from Eq.~\eqref{equ:phase_measure_def} that the SSync measure $S_\mathrm{max} = 0$ in this case of diagonal steady state.

\heading{The steady states of the spin-1 model.}\label{appendix:B}---We show the properties, referred to in the main text, of the steady states of the spin-1 model whose Hamiltonian $H$ and jump operators $L_j$ are defined in Eq.~\eqref{equ:spin1:H} and below. We will begin by reviewing the definition of strong symmetry and its implication on the number of linearly independent steady states. We then identify the steady states using the model's property that all the jump operators are Hermitian. Last, we specify the off-diagonal elements of these steady states in the computation basis. We will also remark that the steady states are diagonal once the whole system except for one spin is traced out: They therefore satisfy the limit-cycle requirement mentioned in the main text.

In general, a \textit{strong symmetry} is generated by an operator that commutes with the Hamiltonian, all the jump operators, and their Hermitian conjugates~\cite{BP12}. As a result, the Hilbert space of the Hamiltonian can be divided into sectors $\mathbf{H} = \oplus_\alpha \mathbf{H}_\alpha$ corresponding to the quantum numbers $\alpha$ of the symmetry: The Hamiltonian and the jump operators then take block-diagonal structures in accordance with this direct sum. Consequently, every operator in $\mathbf{B}_{\alpha, \beta} = \left\{\ket{\psi}\bra{\phi}: \ket{\psi}\in \mathbf{H}_\alpha, \ket{\phi}\in \mathbf{H}_\beta\right\}$ evolves within the same subspace, based on the general master equation given in Eq.~\eqref{equ:def_Liouvillian}. 

Now consider the time evolution within an operator subspace $\mathbf{B}_{\alpha, \alpha}$. The evolution can be viewed as one following a Liouvillian given in Eq.~\eqref{equ:def_Liouvillian} with the Hamiltonian $\I_ \alpha H \I_\alpha$ and the jump operators $\I_\alpha L_\mu \I_\alpha$ for all $\mu$. As is in the main text, $\I_\alpha$ denotes the projector into the Hilbert space sector $\mathbf{H}_\alpha$ and, in terms of an orthonormal basis $\mathcal{B}_\alpha$ spanning $\mathbf{H}_\alpha$, takes the form
\eq{
    \I_\alpha = \sum_{\ket{\psi}\in \mathcal{B}_\alpha} \ket{\psi}\bra{\psi}. \label{equ:projector_alpha}
}
Since a Liouvillian is guaranteed to have a steady state~\cite{Evans77}, one concludes~\cite{BP12} that $\mathbf{B}_{\alpha, \alpha}$ for every $\alpha$ contains at least one steady state. (Note however that one cannot conclude the same for $\mathbf{B}_{\alpha, \beta}$ with $\alpha \neq \beta$, since all the operators in such a subspace are traceless.) Specifically, the $N$ spin-1 model has a strong symmetry generated by the total magnetization $M = \sum_{j=1}^N S_j^z$ and the global spin flip $P = \sum_\mathbf{m} \ket{-\mathbf{m}}\bra{\mathbf{m}}$. As we have discussed in the main text, the symmetry leads to the quantum numbers $-N, \cdots, -1, 0+, 0-, 1, \cdots, N$, so we expect at least $2(N+1)$ linearly independent steady states.

To identify the steady states of the spin-1 model, we note that the model's jump operators $L_j = \sqrt{\gamma} \left(S_j^z\right)^2$ [given below Eq.~\eqref{equ:spin1:H}] are all Hermitian. With this property, we find in every operator subspace $\mathbf{B}_{\alpha, \alpha}$ a steady state $\NESSRDM_\alpha = \I_\alpha / D_\alpha$, where $D_\alpha$ is the dimension of the Hilbert space sector $\mathbf{H}_\alpha$, and $\I_\alpha$ is the projector into it [cf. Eq.~\eqref{equ:projector_alpha}]. Indeed, recall that the Hamiltonian $H$ and all the jump operators in the Liouvillian $\mathcal{L}$ given in Eq.~\eqref{equ:def_Liouvillian} are block-diagonal and thus commute with $\NESSRDM_\alpha \propto \I_\alpha$. The Hermiticity of the jump operators then leads to $\mathcal{L}[\NESSRDM_\alpha] = 0$.

\begin{figure}
    \centering
    \includegraphics[width=0.99\linewidth]{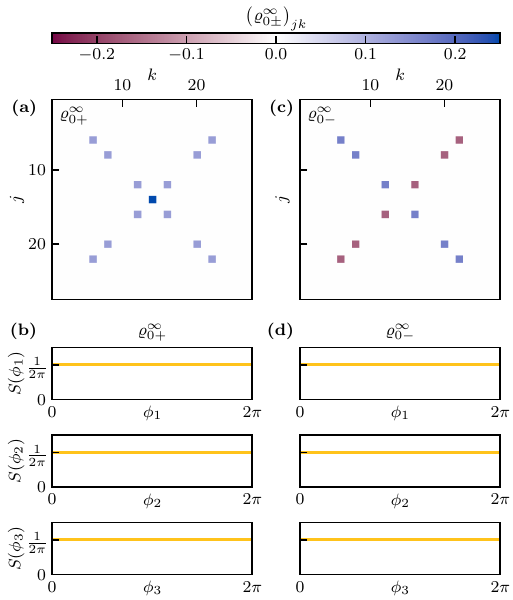}
    \caption{{ (a), (c)} The off-diagonal structure of $\NESSRDM_{0\pm}$, two of the steady states in the $N=3$ spin-1 model, respectively [cf. Eq.~\eqref{equ:spin1:off-diagonals}]. { (b), (d)} The probability distribution $S(\phi_j)$ of the azimuthal spin angle $\phi_j$ of the $j$-th spin ($j = 1, 2, 3$), with the rest of the spins traced out. These distributions are related to the joint distribution $S(\phi_1, \phi_2, \phi_3)$ [defined in Eq.~\eqref{equ:S_phi_spin1}] by, for example, $S(\phi_2) = \int_0^{2\pi} d\phi_1 \int_0^{2\pi} d\phi_3 \,S(\phi_1, \phi_2, \phi_3)$. We have obtained $\NESSRDM_{0\pm}$ by projecting a numerically computed steady state into the $\mathbf{B}_{0+, 0+}$, $\mathbf{B}_{0-, 0-}$ sectors, respectively, at $\omega = J = 2\Delta = \gamma/2 = 1$ [cf. Eq.~\eqref{equ:spin1:H}], and we do not observe change in results upon varying these parameters.
    }
    \label{fig:limit_cycles}
\end{figure}

Next, we examine the structure of the steady states $\NESSRDM_\alpha = \I_\alpha / D_\alpha$ of the spin-1 model, in the computation basis. For $\alpha = \pm 1, \cdots, \pm N$, we can choose the orthonormal basis $\mathcal{B}_\alpha$ spanning the Hilbert space sector $\mathbf{H}_\alpha$ to be $\mathcal{B}_\alpha = \{\ket{\mathbf{m}}: \sum_{j=1}^N m_j = \alpha\}$. We thus see from Eq.~\eqref{equ:projector_alpha} that $\NESSRDM_\alpha$ is diagonal in the computation basis. On the other hand, for $\alpha = 0\pm$ a good basis is $\mathcal{B}_{0\pm} = \{\ket{\psi^\pm_\mathbf{m}}\}$, where we define $\ket{\psi_\mathbf{m}}^\pm = \ket{\mathbf{m}} \pm \ket{\mathbf{-m}}$, up to normalization, for every $\mathbf{m}$ such that $\sum_j m_j = 0$ (except for the case $m_j=0\;\forall j$, where we omit $\ket{\psi_\mathbf{m}^-}$). In this case, the off-diagonal entries in $\NESSRDM_{0\pm}$ are those proportional to $\pm \ket{\mathbf{m}}\bra{-\mathbf{m}}$, respectively, where $\mathbf{m}$ satisfies $\sum_j m_j = 0$ yet $\mathbf{m} \neq \mathbf{0}$. In the main text, we have also discussed the specific example $N=3$, where the off-diagonal entries are, explicitly,
\al{
    &\NESSRDM_{0\pm} - \mathrm{diag}[\NESSRDM_{0\pm}] = \pm \frac{1}{2D_{0\pm}}\left(\ket{1,-1,0}\bra{-1,1,0}\right. \nonumber \\
    &+ \left.\ket{0,1,-1}\bra{0,-1,1} + \ket{-1,0,1}\bra{1,0,-1} + \hc\right),
    \label{equ:spin1:off-diagonals}
}
with $D_{0+} = 4$ and $D_{0-} = 3$. See Fig.~\ref{fig:limit_cycles} for illustration.

Every steady state $\NESSRDM_{\alpha}$ of the spin-1 model is nevertheless diagonal in the computation basis once all the spins except for one are traced out. This clearly holds for $\alpha = \pm1, \cdots, \pm N$ as $\NESSRDM_{\alpha}$ in these cases are diagonal by themselves. To demonstrate this for $\NESS_{0\pm}$, assume without loss of generality that the spins at $j\neq 1$ are traced out. The off-diagonal entries in $\NESS_{0\pm}$, as we have explained above, are located at $\ket{\mathbf{m}}\bra{-\mathbf{m}}$ such that $\sum_{j=1}^N m_j = 0$ yet $\mathbf{m}\neq \mathbf{0}$. However, an off-diagonal entry of the form $\ket{\mathbf{m}}\bra{-\mathbf{m}}$ survives the partial trace only if $m_j = 0$ for all $j\neq 1$, which is impossible given the conditions on $\mathbf{m}$. Therefore, none of the off-diagonal entries remain once the other spins are traced out.

While we have presented analytic arguments for the above structures of the steady states for general $N$ number of spins,  now we illustrate these properties at $N=3$ in Fig.~\ref{fig:limit_cycles}. We focus on the steady states $\NESSRDM_{0\pm}$ that contain off-diagonal entries in the computation basis. The only two steady states that contain off-diagonal entries, namely $\NESSRDM_{0\pm}$, are presented in Figs.~\ref{fig:limit_cycles}\textcolor{vibrant}{a} and~\ref{fig:limit_cycles}\textcolor{vibrant}{c}, respectively. We have elucidated in the main text that these off-diagonal entries contribute to SSync. However, for every single azimuthal angle $\phi_j$, they do not lead to any preference for a particular direction. The latter limit-cycle property is illustrated in Figs.~\ref{fig:limit_cycles}\textcolor{vibrant}{b} and~\ref{fig:limit_cycles}\textcolor{vibrant}{d} for $\NESSRDM_{0\pm}$, respectively, and indicates that each of the synchronizing spins is free on its own. As we have remarked in the main text, this limit-cycle requirement is crucial for establishing synchronization; otherwise, the predictable angular differences among $\phi_j$ may be only a result of individually polarized spins. The steady states $\NESSRDM_{0\pm}$ indeed contain \textit{correlation} among the \textit{free} angles that gives rise to SSync.

\clearpage
\setcounter{section}{0}
\renewcommand{\thesection}{\Roman{section}}

\newpage
\onecolumngrid

\begin{center}
\large{\textbf{Supplemental Material}}
\end{center}

\edef\maineqcnt{\arabic{equation}}
\edef\mainfigcnt{\arabic{figure}}

\renewcommand{\theequation}{S\the\numexpr\value{equation}-\maineqcnt\relax}
\renewcommand{\thefigure}{S\the\numexpr\value{figure}-\mainfigcnt\relax}
\renewcommand{\thesection}{S\Roman{section}}
\renewcommand{\bibnumfmt}[1]{[S#1]}

\newcommand{\maineqref}[1]{\eqref{#1}}
\newcommand{\mainref}[1]{\ref{#1}}

\section{On the steady states of decoupled oscillating coherences}
\label{sec:NESS_of_OSC}

This section concerns the steady states $\sigmaRDM_1$, $\sigmaRDM_2$, and $\NESSRDM_z$ that appear in the discussion in the main text on two decoupled spins each hosting a pair of oscillating coherences. We first show that, for every individual spin of concern, there exists a unitary matrix $U$ acting on it such that $\sigmaRDM_1 = U\OSC$, $\sigmaRDM_2 = \OSC U$ are steady states. Normalization $\tr[\sigmaRDM_1] = \tr[\sigmaRDM_2] = 1$ can be achieved by rescaling $\OSC$ if necessary. We then show that $\NESSRDM_z$, constructed in Eq.~\maineqref{equ:mixture_twosites} for all $z = u e^{i\xi}$ such that $0<u<1$ and $\xi\in\R$, are well-defined steady states of the entire system consisting of the two spins.

The first statement follows largely from Theorem 1 of Ref.~\cite{Buca22}, which states that a Liouvillian $\mathcal{L}$ of the form given in Eq.~\maineqref{equ:def_Liouvillian} has an oscillating coherence $\OSC$ with an eigenvalue $\j \omega$ for some $\omega\in\R$ if and only if 
\eq{
    \OSC = U^\dagger \sigmaRDM \label{equ:polar_decomposition},
}
for a steady state $\sigmaRDM$ and a unitary matrix $U$ that satisfy
\al{
    \sigmaRDM[L_\mu^\dagger, U]  &= 0\quad \forall  \mu \label{equ:osc_if_1}\\
    \sigmaRDM \left(-\j[H, U] + \frac{1}{2}\sum_\mu L_\mu^\dagger[L_\mu, U]   \right)  &= -\j \omega  \sigmaRDM U.
    \label{equ:osc_if_2}
}
Here, we present the above two equations in their Hermitian conjugated forms of the ones listed in Ref.~\cite{Buca22}, for easy future reference. The factor $1/2$ appears because of the different convention we have adopted for the Liouvillian $\mathcal{L}$ in Eq.~\maineqref{equ:def_Liouvillian}. The polar decomposition given in Eq.~\eqref{equ:polar_decomposition} directly shows that $\sigmaRDM = U \OSC$, which we denote by $\sigmaRDM_1$ in the main text, is a steady state. We recall that by \textit{steady state} we refer to a positive semidefinite (hence Hermitian) eigenmode $\NESSRDM$ of $\mathcal{L}$ such that $\mathcal{L}[\NESSRDM] = 0$ and $\tr[\NESSRDM] = 1$: While Ref.~\cite{Buca22} does not explicitly require the latter condition of unit trace, it can be achieved by properly rescaling $\OSC$. Next, we show that $\sigmaRDM_2 = \OSC U$ is a steady state as well. Indeed, note $\sigmaRDM_2 = U^\dagger \sigmaRDM_1 U$ and thus the unitarity of $U$ implies that $\tr[\sigmaRDM_2] = \tr[\sigmaRDM_1] = 1$ and that $\sigmaRDM_2$ is positive semidefinite as $\sigmaRDM_1$ is so. It remains to show that $\mathcal{L}[\sigmaRDM_2] = 0$, which can be done explicitly with Eq.~\maineqref{equ:def_Liouvillian},
\eq{
    \mathcal{L}[\sigmaRDM_2] = -\j [H, U^\dagger \sigmaRDM_1 U] + \sum_\mu\left[L_\mu U^\dagger \sigmaRDM_1 U L_\mu^\dagger  - \frac{1}{2}\left(L_\mu^\dagger L_\mu U^\dagger \sigmaRDM_1 U + U^\dagger \sigmaRDM_1 U L_\mu^\dagger L_\mu\right)\right].
}
We want to make use of $\mathcal{L}[U^\dagger \sigmaRDM_1] = \j \omega U^\dagger \sigmaRDM_1$ and to this end reorganize as follows
\eq{
\mathcal{L}[\sigmaRDM_2] = \mathcal{L}[U^\dagger \sigmaRDM_1]U - \j U^\dagger \sigmaRDM_1[H, U] + \sum_\mu \left( L_\mu U^\dagger \sigmaRDM_1 [U, L_\mu^\dagger] - \frac{1}{2} U^\dagger \sigmaRDM_1 [U, L_\mu^\dagger L_\mu] \right).
}
Note that the first term in the parentheses is zero because of Eq.~\eqref{equ:osc_if_1}, which also allows us to simplify the second term thereof to be $-U^\dagger \sigmaRDM_1 L_\mu^\dagger [U, L_\mu]/2 = U^\dagger \sigmaRDM_1 L_\mu^\dagger [L_\mu, U]/2$. To use the other condition, Eq.~\eqref{equ:osc_if_2}, we pull $U^\dagger \sigmaRDM_1$ to the left and obtain
\eq{
\mathcal{L}[\sigmaRDM_2] = \mathcal{L}[U^\dagger \sigmaRDM_1]U + U^\dagger \sigmaRDM_1 \left(-\j [H, U] + \frac{1}{2}L_\mu^
\dagger [L_\mu, U]\right).
}
Equation~\eqref{equ:osc_if_2} therefore leads to
\eq{
\mathcal{L}[\sigmaRDM_2] = \mathcal{L}[U^\dagger \sigmaRDM_1]U - \j \omega U^\dagger \sigmaRDM_1 U = 0,
}
as desired. This concludes our proof for the existence of steady states of the form $\sigmaRDM_1 = U\OSC$, $\sigmaRDM_2 = \OSC U$, claimed in the main text for every single spin concerned.

The same theorem~\cite{Buca22} is crucial for establishing that $\NESS_z$ as constructed in Eq.~\maineqref{equ:mixture_twosites} of the main text for the two-spin system are well-defined steady states for all $z$. To this end, it suffices to show that the steady states $\rho_\xi$ parametrized by $u=1$ and $\xi\in \R$, namely
\eq{
    \NESSRDM_\xi = \frac{1}{2}\left[\sigmaRDM_{1, \mathrm{A}} \otimes \sigmaRDM_{2, \mathrm{B}} + \sigmaRDM_{2, \mathrm{A}} \otimes \sigmaRDM_{1, \mathrm{B}} + \left(e^{\j \xi}\rho^{\lambda=0}_\mathrm{new} + \hc\right)\right],
}
are well-defined for all $\xi$. If this is true, then for every $z = u e^{\j \xi}$ such that $0 < u < 1$ the state $\NESSRDM_z$ is well-defined because we can write $\NESSRDM_z$ as a probability mixture of two well-defined density matrices,
\eq{
    \NESSRDM_z = \frac{1+u}{2} \NESSRDM_{\xi} + \frac{1-u}{2} \NESSRDM_{\pi + \xi}.
}
Note that indeed we can interpret $(1\pm u)/2$ as probabilities because the two numbers are positive and sum to unity. 

We now show that $\NESSRDM_\xi$ is well-defined for every $\xi\in \R$. That $\tr[\NESSRDM_\xi] = 1$ follows directly from the fact that $\tr[M_\mathrm{A} \otimes M_\mathrm{B}] = \tr[M_\mathrm{A}] \tr[M_\mathrm{B}]$ for all matrices $M_\mathrm{A}, M_\mathrm{B}$; also note $\sigmaRDM_{\alpha, j}$ for all $\alpha = 1, 2$ and $j=\mathrm{A}, \mathrm{B}$ are well-defined steady states with unit trace, and $\OSC_j$ as eigenmodes of the Liouvillian $\mathcal{L}_j$ with eigenvalues $\j \omega \neq 0$ all have zero trace. It remains to show that $\NESSRDM_\xi$ is positive semidefinite for every $\xi$. Indeed, $\NESSRDM_\xi = V_\xi^\dagger (\sigmaRDM_{1, \mathrm{A}}\otimes \sigmaRDM_{1, \mathrm{B}}) V_\xi$ where
\eq{
V_\xi = U_\mathrm{A} \otimes \I_\mathrm{B} + e^{\j\xi}  \I_\mathrm{A} \otimes U_\mathrm{B}.
}
Here, as is in the main text, the subscripts A, B specify the spin that each single-spin operator acts on. Consequently, supposing a decomposition of the steady state $\sigmaRDM_{1, \mathrm{A}} \otimes \sigmaRDM_{1, \mathrm{B}} = \sum_n p_n \ket{\psi_n}\bra{\psi_n}$ in terms of orthonormal states $\ket{\psi_n}$ and probabilities $p_n > 0$, we find $\NESSRDM_\xi = \sum_n p_n \ket{\phi_n}\bra{\phi_n}$ in terms of the states $\ket{\phi_n} = V^\dagger_\xi \ket{\psi_n}$. It therefore follows that, for every state $\ket{\xi}$ of the entire system, $\bra{\xi} \NESSRDM_\xi \ket{\xi} = \sum_n p_n \left|\braket{\phi_n | \xi }\right|^2 \ge 0$, so $\NESSRDM_\xi$ is positive semidefinite. This confirms that $\NESSRDM_\xi$ are well-defined steady states and therefore concludes the proof.

\section{The Generalized $N$ Spin-$s$ Model}
\label{sec:spin-S}
We discuss a generalization of the dissipative spin-1 model described in the main text to an $N$ spin-$s$ model, whose Hamiltonian and the jump operators take the same form as given in Eq.~\maineqref{equ:spin1:H} and below. We require $s\ge 1$ as otherwise the jump operators $L_j = \sqrt{\gamma} \left(S_j^z\right)^2$ are trivial. The computation basis is now spanned by $\mathbf{m} = (m_1, \cdots, m_N)$, with $-s \leq m_j \leq s$. Much of the analysis of the spin-1 model directly carries over because the $\mathrm{U}(1)\rtimes\Z_2$ strong symmetry still holds, generated by the total magnetization $M = \sum_j S^z_j$ and the global inversion $P = \bigotimes_j \sum_m(\ket{m}_j\bra{m}_j)$, with $PM=-MP$. 

The strong symmetry divides the Hilbert space into orthogonal sectors, such that a pure initial state in a given sector does not evolve out of it---we expect one steady state per sector (cf. Appendix~\mainref{appendix:B}). The quantum numbers of the total magnetization $M$ take values from $-Ns, \cdots Ns$. If $Ns$ is an integer, then the $M=0$ sector exists, which is further divided by the global inversion $P$ into sectors $0\pm$, corresponding to the quantum numbers $\pm 1$, respectively, as is the case in the main text. Otherwise, there is no $M=0$ sector, and the total magnetization $M$ itself serves as the quantum number.

In the following, we elaborate in Section~\ref{sec:spin-S:steady_states} on the structure of the steady states corresponding to these Hilbert space sectors. In Section~\ref{sec:spin-S:DSync}, we establish that the model exhibits DSync if and only if $s$ is an integer. In Section~\ref{sec:spin-S:SSync}, we show that, just like the spin-1 model in the main text, the spin-$s$ model with integer $s$ exhibits signatures of SSync. We also comment in Section~\ref{sec:spin-S:SSync} that for half-integer $s$, where the model has no DSync, the system does not show SSync either. In summary, our results suggest that DSync comes hand in hand with SSync in this strongly coupled spin-$s$ model.

\subsection{The structure of steady states}
\label{sec:spin-S:steady_states}

Because the model still has only Hermitian jump operators [cf. below Eq.~\maineqref{equ:spin1:H}, also explained in Appendix~\mainref{appendix:B}], the steady state for each sector $\alpha$ takes the form $\NESSRDM_\alpha \propto \I_\alpha$. The projector $\I_\alpha$ onto the Hilbert space sector $\alpha$ is diagonal in the computation basis except when $\alpha = 0\pm$, which as we have analyzed above is possible only for integer $Ns$. For the latter cases $\alpha = 0\pm$, the off-diagonal elements are proportional to $\ket{\mathbf{m}}\bra{-\mathbf{m}}$, with $\sum_j m_j = 0$ while $\mathbf{m} \neq \mathbf{0}$, that nevertheless do not survive when all the spins except for one are traced out. These conclusions on the structures of the steady states $\NESSRDM_\alpha$ follow from the same analysis in Appendix~\mainref{appendix:B}.

\subsection{DSync for integer $s$ only}
\label{sec:spin-S:DSync}

The oscillating coherences follow the same formal expression as those of the spin-1 model do, because it directly results from, as we have derived in the main text, the algebraic relations among the global inversion $P$, the total magnetization $M$, and the Hamiltonian $H$. In particular, the purely imaginary eigenvalues are $2\j \omega M$, for all the total magnetization quantum numbers $M$ such that $M\neq 0$, and the corresponding eigenmodes are $\OSC_M = P\I_M$.

For a single-site operator $O_{m, j}$ (in terms of transverse spin operators in the $x$-$y$ plane, cf. main text) on spin $j$ to exhibit DSync, it must overlap with an oscillating coherence $\OSC_m$, namely $\tr [O_{m, j}\OSC_m]\neq 0$. In other words, it must hybridize two states that differ only at one spin yet have opposite polarizations at all spins. This is impossible for half-integer spins, where such hybridization can only be achieved by a global operator. For the spin $s$ being integer, this is possible only if $O_{m, j}$ hybridizes the $\ket{-m}_j \otimes \ket{0\cdots}$, $\ket{m}_j \otimes \ket{0\cdots}$ states, which constrains $m = 1, \cdots, s$. Furthermore, we construct $O_{m, j} = (S^x_j)^{2m}$; indeed the family of operators $O^\phi_{m, j} = e^{\j \phi S^z_j} O_{m, j} e^{-\j \phi S^z_j}$, which are obtained with this operator rotated by the azimuthal angle $\phi$, exhibits $\tr[O^\phi_{m, j} \OSC_m] \propto e^{2\j m \phi}$. On the other hand, the limit-cycle condition is satisfied as $\tr[O^\phi_{m, j} \NESSRDM_\alpha]$, for every sector label $\alpha$,  picks out  only the diagonal entries of $O^\phi_{m, j}$ and thus has no $\phi$ dependence, given the  structures of $\NESSRDM_\alpha$ in Sec.~\ref{sec:spin-S:steady_states}. These results show that DSync happens only in the integer-$s$ cases.

\subsection{DSync implies SSync}
\label{sec:spin-S:SSync}
The operators $O_{m, j} = (S^x_j)^{2m}$ that probe DSync at integer $s$, where $m = 1, \cdots, s$, exhibit nontrivial correlations in the steady states $\NESSRDM_{0\pm}$ of the $0\pm$ sectors, which exist for integer $s$ (cf. Sec.~\ref{sec:spin-S:steady_states}). Indeed, we can compute with the knowledge of the off-diagonal elements described in Sec.~\ref{sec:spin-S:steady_states} that the steady states $\NESSRDM_{0\pm} \propto \I_{0\pm}$ show correlations $\tr[O^{\phi_j}_{m, j} O^{\phi_k}_{m, k}\NESSRDM_{0\pm}] \propto \cos[2m(\phi_j - \phi_k)]$, for every pair of spins $j$, $k$. Such two-point correlations suggest a statistical preference for the azimuthal angles $\phi_j$, $\phi_k$ to align or anti-align.

These correlations between the azimuthal angles satisfy our condition (ii) in the main text for the generically perturbed model to feature SSync. Condition (i) of the limit-cycle requirement is also satisfied for the same reason elucidated in the main text for the special spin-1 model: For every $\alpha$, the expectation values of the single-site spin operator, $\tr[O_{m, j}^\phi \NESSRDM_\alpha]$, are only contributed by the diagonal entries of $O^\phi_{m, j}$ (since the states $\NESSRDM_\alpha$ with all the spin but one traced out are diagonal, cf. Sec.~\ref{sec:spin-S:steady_states}) and thus do not show dependence on the azimuthal angle $\phi$. These two conditions being satisfied shows that, under generic perturbations, the spin-$s$ model is left with a single steady state that exhibits SSync.

One may wonder whether these signatures of SSync exist even if the model does not feature DSync, i.e., if the spin $s$ is a half integer. For half-integer $s$ and odd $N$, the Hilbert space does not have a $0\pm$ sector, so every steady state $\NESSRDM_\alpha$, now labeled by $\alpha = -Ns, \cdots, -1/2, 1/2, \cdots, Ns$, is diagonal in the computation basis (cf. Sec.~\ref{sec:spin-S:steady_states}). This suggests that any operator family of the form $Q^{\phi_j}_{n, j} = e^{\j\phi_j S^z_j} (S^x_j)^n e^{-\j \phi_j S^z_j}\; \forall n = 1, \cdots, 2S$, which are the $n$-th power of the transverse spin operator rotated by the local azimuthal angle $\phi_j$, does not show nontrivial $l$-point correlations (for all $l = 1, \cdots N$) that depend on the azimuthal angles. For half-integer $s$ and even $N$, the above arguments apply as well to all the steady states $\NESSRDM_M$ with $M\neq 0$, but we need to examine the non-diagonal steady states $\NESSRDM_{0\pm}$ now that the model has $0\pm$ sectors, $Ns$ being an integer. Notably, the off-diagonal elements still take the form of $\ket{\mathbf{m}}\bra{-\mathbf{m}}$ with $\sum_j m_j = 0$ (cf. Sec.~\ref{sec:spin-S:steady_states}), which can only be detected by global operators as $m_j = -m_j$ is impossible for the half-integer spin $s$; in other words, tracing out any subsystem (of arbitrary, nonzero length) will eliminate these off-diagonal elements. As a result, the operators $Q^{\phi_j}_{n, j}$ have trivial $l$-point correlation functions for all $l < N$. We therefore conclude that the system features no meaningful SSync for half-integer $s$, in which case there is no DSync, either.

\section{Details on Numerical Analysis}
\label{sec:numerics}
In this Section, we provide additional information on our numerical approaches and results. In Sec.~\ref{sec:numerics:sampling}, we specify the method taken to sample the generic perturbations in verifying our analytical predictions for the three spin-1 model in the main text. We will apply the same method to a spin-1/2 model later in Sec.~\ref{appendix:spin1/2}. In Sec.~\ref{sec:numerics:chi}, we motivate the function $\chi$ defined in Eq.~\maineqref{equ:chi_phase_diff} of the main text and show its key properties, which allow us to characterize the configuration of the three azimuthal angles in the spin-1 model. In Sec.~\ref{sec:numerics:quantitative}, we present more quantitative data of the same model and show that the $(0, \pi)$ and $\pi/3$ types of steady-state synchronization, whose occurrence we have predicted analytically and verified numerically with qualitative methods in the main text, occur roughly at the same rates for our ensemble of randomness.

\subsection{Sampling generic perturbations}
\label{sec:numerics:sampling}

We elaborate on how we sample the generic forms of perturbations $\mathcal{L}_{(1)}$ from an ensemble of random Liouvillians that is proposed in Ref.~\cite{Denisov19}. The system under perturbation is then governed by the Liouvillian $\mathcal{L} + \eta \mathcal{L}_{(1)}$, where $\mathcal{L}$ describes the unperturbed system and $\eta$ tunes the perturbation strength. The perturbative Liouvillian $\mathcal{L}_{(1)}$ consists of two parts,
\eq{
\mathcal{L}_{(1)}[\varrho] = \j \zeta_\mathrm{coh} (\varrho H_{(1)} - H_{(1)}  \varrho) + \zeta_\mathrm{diss} \mathcal{L}^\mathrm{diss}_{(1)}[\varrho].
}
Here, the first of the two terms constitutes the coherent part, evolving the system by a random Hamiltonian $H_{(1)}$ acting on a Hilbert space of dimension $D_\mathbf{H}$, and the second term is a purely dissipative part given by a dissipator $\mathcal{L}^\mathrm{diss}_{(1)}$. Their relative contributions to the overall perturbative Liouvillian $\mathcal{L}_{(1)}$ are tuned by the real numbers $\zeta_\mathrm{coh}$, $\zeta_\mathrm{diss}$.  While Ref.~\cite{Denisov19} has investigated different sets of the two parameters, for our purpose we take $\zeta_\mathrm{coh} = \zeta_\mathrm{diss} = 1$ unless noted otherwise; we expect this choice to represent the generic situation. We sample $H_{(1)}$ and $\mathcal{L}^\mathrm{diss}_{(1)}$ independently based on the procedures described in Ref.~\cite{Denisov19}, which we also summarize as follows. 

We sample the Hamiltonian $H_{(1)}$ from a Gaussian unitary ensemble of unit variance. Following the standard procedure (see for example in Ref.~\cite{AGZ09}), we first sample a random $D_\mathbf{H}$-by-$D_\mathbf{H}$ matrix $M$, the real and imaginary parts of whose entries are i.i.d. Gaussian random variables of mean $0$ and variance $1$. We then take $H_{(1)} = \left(M + M^\dagger\right) / \sqrt{2}$.

To sample the dissipator $\mathcal{L}^\mathrm{diss}_{(1)}$ following Ref.~\cite{Denisov19}, we start by generating a $d$-by-$d$ matrix $G$ with real and imaginary parts of the entries being i.i.d. Gaussian random variables of mean $0$ and variance $1$; here $d = D_\mathbf{H}^2 - 1$. The general form of the dissipator is then given in terms of a properly normalized matrix $K = d GG^\dagger/\tr[GG^\dagger]$,
\eq{
\mathcal{L}^\mathrm{diss}_{(1)}[\varrho] = \sum_{m, n=1}^{d} K_{mn} \left[F_n \varrho F_m^\dagger - \frac{1}{2}\left\{F_m^\dagger F_n, \varrho\right\} \right], \label{equ:sampling_LD1}
}
for traceless matrices $F_n$ satisfying $\tr[F_m F_n^\dagger] = \delta_{m, n}$. In practice, we pick the $D_\mathbf{H}$-by-$D_\mathbf{H}$ matrices $F_n$, where $n = 1, \cdots, d$, so that their entries (indexed by $j, k$) are
\eq{
    \left(F_{(a-1)D_\mathbf{H} + b}\right)_{j, k} =
    \begin{cases}
        \delta_{j, a} \delta_{k, b} & a\neq b \\ 
        \delta_{j, a} \delta_{k, a} - \delta_{j, a+1} \delta_{k, a+1} & a = b
    \end{cases},
}
for $a, b = 1, \cdots, D_\mathbf{H}$ except for $a=b=D_\mathbf{H}$.

While Eq.~\eqref{equ:sampling_LD1} is sufficient for describing our sampling approach for the perturbative dissipator $\mathcal{L}^\mathrm{diss}_{(1)}$, we remark in the following (as has been pointed out in Ref.~\cite{Denisov19}) that it can be written in the standard form of Lindblad master equations [cf. Eq.~\maineqref{equ:def_Liouvillian}]. Indeed, after diagonalizing the Hermitian positive semidefinite matrix $K$ as $K_{\sigma, \tau} = \sum_\mu \kappa_\mu V_{\mu,\sigma}^* V_{\mu, \tau}$ in terms of a $d$-by-$d$ unitary matrix $V$ and eigenvalues $\kappa_\mu \geq 0$, we can construct the jump operators $L_\mu = \sum_{\nu} \sqrt{\kappa_\mu} V_{\mu, \nu} F_\nu$ for $\mu = 1, \cdots, d$. Then, Eq.~\eqref{equ:sampling_LD1} is rewritten as
\eq{
    \mathcal{L}^\mathrm{diss}_{(1)}[\varrho] = \sum_{\mu=1}^{D_\mathbf{H}^2-1} \left(L_\mu \varrho L_\mu^\dagger - \frac{1}{2} \left\{L_\mu^\dagger L_\mu,  \varrho\right\}\right),
}
in the desired form.

\subsection{The characterization function $\chi$ of differences among three angles}
\label{sec:numerics:chi}

We comment on and derive the properties of the function $\chi(\phi_1, \phi_2, \phi_3)$ defined in Eq.~\maineqref{equ:chi_phase_diff} of the main text. To this end, we consider the function $l = \sqrt{\chi}$, which reads explicitly
\eq{
l(\phi_1, \phi_2, \phi_3) = \frac{3}{4\pi} \min_{j\in\{0, 1, 2\}} \sum_i \argdist(\phi_i, \overline{\phi}_j),
}
as a sum of the angular distances [on a circle, defined accompanying Eq.~\maineqref{equ:chi_phase_diff}] from each angle to an optimal (in the sense that the sum is minimized) average angle $\overline{\phi}_j = (2\pi j + \sum_k \phi_k)/3$. We always equate $\phi \equiv \phi + 2\pi n$ for every angle $\phi$ and integer $n$, so the average angle is defined up to $2\pi/3$. While equivalent representatives of the angles $\phi_1, \phi_2, \phi_3$ may affect $\overline{\phi}_j$ for a specific $j$, the set $\left\{\overline{\phi}_j\right\}_{j = 0, 1, 2}$ is invariant. The function $l$ and thus $\chi$ are therefore well-defined functions of the three angles.

We further note that $\chi$, $l$ are invariant under global phase shifts mapping $\phi_j \mapsto \phi_j + \theta \; \forall j$ by a constant $\theta$; indeed, these shifts also map the average angles in the way $\overline{\phi}_k \mapsto \overline{\phi}_k + \theta\; \forall k$, so the angular distances, $\mathrm{argdist}(\phi_j, \overline{\phi}_k)$, are unchanged. Therefore, $\chi$, $l$ are sensitive to only the differences among the angles $\phi_i$ and thus suitable for characterizing synchronization. In particular, $\chi(\phi_1, \phi_2, \phi_3) = \chi(\phi_1', \phi_2', 0)$ and $l(\phi_1, \phi_2, \phi_3) = l(\phi_1', \phi_2', 0)$ in terms of the differences $\phi_j' = \phi_j - \phi_3$, for $j=1, 2$,  defined in the main text.

\begin{figure}
    \centering
    \includegraphics[width=0.5\linewidth]{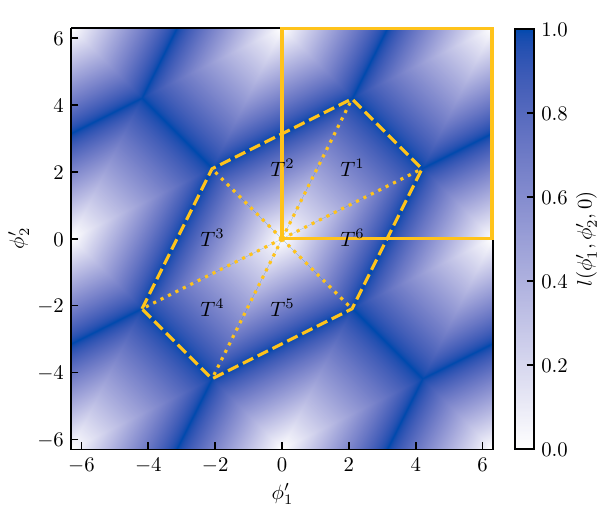}
    \caption{The $l=\sqrt{\chi}$ values of the angular differences $\phi_1', \phi_2'$, each ranging from $-2\pi$ to $2\pi$. The region $\Omega_d$ at $d=1$ is enclosed in dashed lines and has an area equal to (as can be seen by shifting $\phi_1'$ or $\phi_2'$ by $2\pi$) that of the square $[0, 2\pi]\times [0, 2\pi]$, enclosed by the yellow solid lines. The dotted lines illustrate, at $d=1$, the separation of the region $\Omega_{d}$ into the triangles $T^a_d$.}
    \label{fig:chi3}
\end{figure}

The $\chi$ function bears another useful property, namely that $\chi(\phi_1, \phi_2, \phi_3)$ is uniformly distributed in $[0, 1]$ if the angles $\phi_1$, $\phi_2$, $\phi_3$ are uniformly i.i.d. 
In the main text, we have used this property to argue for a non-synchronization regime. To show this property, it suffices to show that $l(\phi_1', \phi_2')$, which we use to denote $l(\phi_1', \phi_2', 0)=\sqrt{\chi(\phi_1', \phi_2', 0)}$, takes values in $[0, 1]$ and follows the distribution $f(l) = 2l$ given uniformly i.i.d. $\phi_1$, $\phi_2$, $\phi_3$, i.e., uniformly i.i.d. angular differences $\phi_1', \phi_2'$.

We start by considering the cumulative distribution $F(l) = \int_0^{l} f(l') dl'$. We claim that, given $d\in (0, 1]$, the following region $\Omega_d\subset \left\{(\phi_1', \phi_2'): \phi_1', \phi_2'\in [-2\pi, 2\pi]\right\}$ given by $ \Omega_d= \bigcup_{a=1}^6 \overline{T^a_d}$ satisfies $l(\Omega_d)< d$ and $l(\partial \Omega_d)=d$, where $T^a_d$ are disjoint triangles,
\al{
    T^1_d &= \left\{0<\phi_1' + \phi_2' < 2\pi d, \phi_1' - 2\phi_2'<0, 2 \phi_1' - \phi_2' > 0\right\} \nonumber \\
    T^2_d &= \left\{0<2\phi_2' - \phi_1' < 2\pi d, 2\phi_1' - \phi_2' < 0, \phi_1' + \phi_2' > 0\right\} \nonumber \\
    T^3_d &= \left\{0<\phi_2' - \phi_1' < 2\pi d, \phi_1'+\phi_2'<0, \phi_1'-2\phi_2' < 0\right\} \nonumber \\
    T_d^4 &= -T_d^1, \quad T_d^5 = -T_d^2, \quad T_d^6 = -T_d^3, \label{equ:Ts}
}
and $\overline{T^a_d}$ denotes the closure of $T^a_d$; we draw $T^a_d$ at $d=1$ in Fig.~\ref{fig:chi3}. Indeed, one can check for every $\phi_1', \phi_2'$ in, e.g., $T_d^1$ that the average angle $\overline{\phi}_j = (2\pi j + \phi_1'+\phi_2')/3$ achieving $l(\phi_1', \phi_2', 0)$ has $j=0$, that
\eq{
    \argdist(\phi_1', \overline{\phi}_0) = \frac{2\phi_1'-\phi_2'}{3}, \quad
    \argdist(\phi_2', \overline{\phi}_0) = \frac{2\phi_2'-\phi_1'}{3}, \quad
    \argdist(0, \overline{\phi}_0) = \frac{\phi_1'+\phi_2'}{3},
}
and therefore that $l(\phi_1', \phi_2')=(\phi_1'+\phi_2')/2\pi<d$. Moreover, $(\phi_1', \phi_2') \in \partial T_d^1$ is in $\partial \Omega_d$ if and only if $\phi_1'+\phi_2'=2\pi d$, so we can conclude for such $\phi_1', \phi_2'$ that $l(\phi_1', \phi_2') = d$. In general, for every $T^a_d$ listed above we can specify the average angle $\overline{\phi}_j$ that is relevant for computing $l$ and thus straightforwardly verify the similar properties. 

We then conclude based on Eq.~\eqref{equ:Ts} and similar equations for all the six triangles $T^a_d$ that the area of $\Omega_d$ is proportional to $d^2$. We furthermore specify this area to be $(2\pi d)^2$ by noting that the area at $d=1$ equals that of a square of length $2\pi$ (cf. Fig.~\ref{fig:chi3}). Since the latter square already contains all the equivalent values of $\phi_1', \phi_2'$, we see that $l$ takes values from $[0, 1]$. Additionally following from the area calculation, for uniformly i.i.d. $\phi_1'$, $\phi_2'$ we have the cumulative distribution $F(l) = (2\pi l)^2/(2\pi)^2 = l^2$ and thus the distribution $f(l) = \mathrm{d} F(l)/\mathrm{d}l = 2l$ as desired.

\subsection{Rates of occurrence of the two types of steady-state synchronization}
\label{sec:numerics:quantitative}

In the main text, we have identified two types of steady-state synchronization in the perturbed three spin-1 model, which are dubbed $(0, \pi)$ and $\pi/3$ types based on the pattern of the likely angular differences. While we have qualitatively verified the occurrence of both types based on the numerical data presented in Fig.~\mainref{fig:varying_eta}\textcolor{vibrant}{a}, we now investigate their occurrence rates quantitatively. Our results indicate that, for the ensemble of randomness of our concern, both types happen at about the same rates. This supports the conclusion in the main text that the $(0, \pi)$ and $\pi/3$ types of steady-state synchronization are both natural to occur under generic perturbations.

The distribution of $\chi$ presented in Fig.~\mainref{fig:varying_eta}\textcolor{vibrant}{a} signifies the probable angular configurations in the steady state and thus contains information on the occurrence rates of interest. Recall from the main text that the distribution of $\chi$ is derived from the distribution $S_\mathrm{d}(\phi_1', \phi_2')$ [whose explicit definition is given in Appendix~\ref{appendix:A} of the main text, Eq.~\maineqref{equ:phase_diff_distribution_def}] of angular differences $\phi_1' = \phi_1 - \phi_3, \phi_2' = \phi_2 - \phi_3$. In particular, we first obtain the distribution of $\chi$ for each single sample of randomness: Every $(\phi_1', \phi_2')$ such that $S_\mathrm{d}(\phi_1', \phi_2') > r \max S_\mathrm{d}$, with $r = 0.95$, contributes a value $\chi(\phi_1', \phi_2', 0)$ weighted by the probability $S_\mathrm{d}(\phi_1', \phi_2')$. We normalize this distribution such that the total probability equals one. In the ideal scenario where $r \rightarrow 1$ and $\eta \rightarrow 0$, we expect the single-sample distribution to take one of the two forms plotted in Figs.~\mainref{fig:S_phis}\textcolor{vibrant}{c} and~\mainref{fig:S_phis}\textcolor{vibrant}{f}, corresponding to the distributions $S_\mathrm{d}$ presented in Figs.~\mainref{fig:S_phis}\textcolor{vibrant}{a} and~\mainref{fig:S_phis}\textcolor{vibrant}{d}, in the cases of $(0, \pi)$ and $\pi/3$ types of synchronization, respectively. At nonzero perturbation strengths $\eta$, we do not assume \textit{a priori} that every sample belongs to either of the two cases, but nevertheless we can look for the features of Figs.~\mainref{fig:S_phis}\textcolor{vibrant}{c} and~\mainref{fig:S_phis}\textcolor{vibrant}{f} for all the samples combined. We then average the aforementioned distribution of $\chi$ over the whole ensemble of generic perturbations. The results are presented in Fig.~\mainref{fig:varying_eta}\textcolor{vibrant}{a}.

As is remarked in the main text, we have chosen $r = 0.95$ slightly below one in order to avoid potential artifacts, as $S_\mathrm{d}$ may contain several local maxima with very similar numerical values (cf. the unperturbed, $\eta = 0$, cases in Figs.~\mainref{fig:S_phis}\textcolor{vibrant}{a} and~\mainref{fig:S_phis}\textcolor{vibrant}{d}). Consequently, for a single realization of randomness we expect this  distribution of $\chi$ to feature smeared-out peaks in place of those presented in either Fig.~\mainref{fig:S_phis}\textcolor{vibrant}{c} or Fig.~\mainref{fig:S_phis}\textcolor{vibrant}{f}. In the ensemble-averaged distribution, we therefore expect three smeared-out peaks combining the features in Figs.~\mainref{fig:S_phis}\textcolor{vibrant}{c} and~\mainref{fig:S_phis}\textcolor{vibrant}{f}. In the main text, we have reported that these peaks indeed appear in Fig.~\mainref{fig:varying_eta}\textcolor{vibrant}{a} at perturbation strengths $\eta \lesssim 10^{-3}\omega$, which we have identified to be the synchronizing regime.

\begin{figure}
    \centering
    \includegraphics[width=0.5\linewidth]{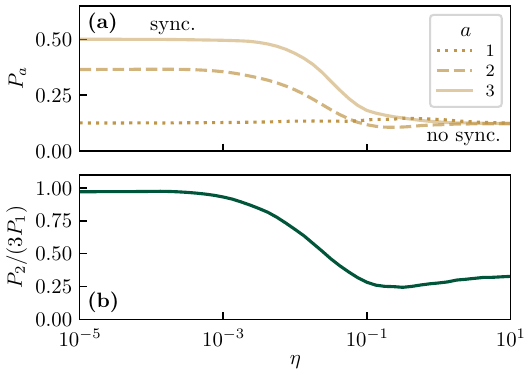}
    \caption{Quantitative analysis for the results presented in Fig.~\mainref{fig:varying_eta}\textcolor{vibrant}{a} of the main text, at various perturbation strengths $\eta$. {\bf (a)} The probability $P_a$ that $\chi$, whose averaged distribution over the whole ensemble of randomness is plotted in Fig.~\mainref{fig:varying_eta}\textcolor{vibrant}{a}, falls within the three same-sized intervals $\mathcal{I}_a$, $a=1, 2, 3$, that enclose the three peaks of the same distribution in the $\eta\ll\omega$ limit, respectively. Here, $\mathcal{I}_1 = [0, 2/16]$, $\mathcal{I}_2 = [3/16, 5/16]$, and $\mathcal{I}_3 = [14/16, 16/16]$. {\bf (b)}~The ratio $P_2/(3P_3)$, which in the $\eta\ll\omega$ limit equals the ratio $\kappa$ of $\pi/3$- to $(0, \pi)$-type of synchronization. }
    \label{fig:quantitative_probs}
\end{figure}

Now we find the ratio $\kappa$ between the occurrence rates of the $(0, \pi)$ and $\pi/3$ types of synchronization in this regime. To this end, we obtain the probability $P_a$ for $\chi$ to fall in same-sized intervals $\mathcal{I}_a$, $a = 1, 2, 3$, each enclosing one of the three peaks in Fig.~\mainref{fig:varying_eta}\textcolor{vibrant}{a}. In particular, we pick $\mathcal{I}_1 = [0, 2/16]$, $\mathcal{I}_2 = [3/16, 5/16]$, and $\mathcal{I}_3 = [14/16, 16/16]$. Note that $\kappa = P_2 / (3P_1)$ in this synchronizing regime; refer to Figs.~\mainref{fig:S_phis}\textcolor{vibrant}{c} and~\mainref{fig:S_phis}\textcolor{vibrant}{f} for the contribution to $P_a$ from the $(0, \pi)$- and $\pi/3$-type samples, respectively. We plot $P_a$ in Fig.~\ref{fig:quantitative_probs}\textcolor{vibrant}{a} and $P_2 / (3P_1)$ in Fig.~\ref{fig:quantitative_probs}\textcolor{vibrant}{b}, for varying $\eta$. In the synchronization regime $\eta /\omega\lesssim 10^{-3}$, we observe $\kappa \approx 1$, indicating that the $(0, \pi)$ and $\pi/3$ types of steady-state synchronization occur at similar rates for the ensemble of perturbations considered.

Additionally, we note that the probabilities $P_a$, for all $a = 1, 2, 3$, converge in the regime $\eta \gtrsim 10^{-1}$. This suggests no fixed angular differences, quantitatively confirming our claim in the main text based on Fig.~\mainref{fig:varying_eta}\textcolor{vibrant}{a} that  synchronization is destroyed in this regime of larger perturbation strengths.

\section{A spin-$1/2$ model of dynamical synchronization}
\label{appendix:spin1/2}
We consider a model of two spin-$1/2$s that is simplified from the one reported to feature dynamical synchronization in Ref.~\cite{Buca22}. The system is governed by the Hamiltonian
\eq{
    H = J \left(\sigma_1^+ \sigma_2^- + \hc \right) + \Delta \sigma_1^z \sigma_2^z + B(\sigma_1^z + \sigma_2^z), \label{equ:spin1/2:H}
}
and jump operators
\eq{
    L_- = \gamma\left(\sigma_1^- + \sigma_2^- \right). \label{equ:spin1/2:L}
}
Here, $\sigma^\pm_j = \left(\sigma^x_j + i\sigma^y_j\right)/2$ in terms of the Pauli operators acting on the spins $j = 1, 2$.

We note a strong inversion symmetry $P$ that is helpful for analyzing the system, given by
\eq{
    P = \ket{\uparrow \uparrow}\bra{\downarrow\downarrow}
    + \ket{\downarrow \downarrow}\bra{\uparrow\uparrow}
    + \ket{\uparrow\downarrow} \bra{\downarrow\uparrow}
    + \ket{\downarrow\uparrow} \bra{\uparrow\downarrow},
}
with eigenvalues $P = \pm 1$. The operator $P$ generates a strong symmetry (defined in Ref.~\cite{BP12}, also briefly reviewed in Appendix~\mainref{appendix:B}) because it satisfies $[H, P] = [P, L_-] =  [P, L_-^\dagger] = 0$. It therefore preserves the dynamics within each of the two Hilbert spaces corresponding to the eigenvalues $\pm1$,
\al{
    \mathbf{H}_+ &= \Span\left\{\ket{\uparrow\uparrow},\; \frac{1}{\sqrt{2}}\left(\ket{\uparrow\downarrow} + \ket{\downarrow\uparrow}\right),\; \ket{v_+} = \ket{\downarrow\downarrow}\right\} \nonumber \\
    \mathbf{H}_- &= \Span\left\{\ket{v_-} = \frac{1}{\sqrt{2}}\left(\ket{\uparrow\downarrow} - \ket{\downarrow\uparrow}\right)\right\},
}
respectively. Here, the states $\ket{v_\pm}$ we have defined are in fact the dark states, as $\NESSRDM_+ = \ket{v_+}\bra{v_+}$ and $\NESSRDM_- = \ket{v_-}\bra{v_-}$ are the steady state of the $\mathbf{B_{++}}$ and $\mathbf{B_{--}}$ operator subspaces, respectively; note that $\ket{v_\pm}$ are eigenstates of the Hamiltonian $H$ with energies $\Delta - 2B$, $-2J - \Delta$, and that $L_-\ket{v_+} = L_-\ket{v_-} = 0$. Following the notations introduced in Appendix~\mainref{appendix:B}, we have defined $\mathbf{B}_{\alpha, \beta} = \left\{\ket{\psi}\bra{\phi}: \ket{\psi}\in \mathbf{H}_\alpha, \ket{\phi}\in \mathbf{H}_\beta\right\}$.

The above analysis implies that the system possesses an oscillating coherence $\OSC = \ket{v_+} \bra{v_-}$ at frequency $\omega = 2(J - \Delta + B)$. It gives rise to DSync (cf. main text) that is probed by the operator $O = \sigma^x$. In particular, the azimuthally rotated operator $O^{\phi_j}_j = e^{\j \phi_j \sigma^z_j/2} \sigma^x_j e^{-\j \phi_j \sigma^z_j/2}$ on spin $j$ overlaps with the oscillating coherence $\OSC$ as 
\al{
\tr[O_1^{\phi_1} \OSC] &= e^{\j \phi_1}/\sqrt{2} \nonumber \\  \tr[O_2^{\phi_2} \OSC] &= e^{\j (\phi_2-\pi)}/\sqrt{2}.
}
According to our general description of DSync detailed in the main text, the expectation values of these observables therefore suggest that, given generic initial conditions, the azimuthal spin angles $\phi_1$ and $\phi_2$ oscillate at the same frequency $\omega$ and fixed phase difference $\pi$ (in other words, they anti-align). This picture agrees with the one introduced in Ref.~\cite{Buca19}, which has referred to it as anti-synchronization. Furthermore, we note that the limit-cycle conditions are strictly satisfied because $\tr[O_1^{\phi_1} \NESSRDM_\alpha]$ and $\tr[O_2^{\phi_2} \NESSRDM_\alpha]$ for all the steady states labeled by $\alpha = +, -$ are $\phi_1$-, $\phi_2$-independent, respectively; indeed, both the two steady states reduced to one spin (with the other spin traced out) are diagonal. We have therefore verified that the anti-synchronization phenomenon studied in Ref.~\cite{Buca19} fits in our description of DSync.

We now elucidate the signatures of SSync in the model by examining the two conditions listed in the main text. Condition (i), the limit-cycle requirement, is satisfied as we have just established that both of the two steady states $\NESSRDM_\pm$ are diagonal (and hence describe free azimuthal angles) when one of the two spins is traced out. Condition (ii) is satisfied because $\NESSRDM_-$, one of the two steady states, exhibits nontrivial correlations detected by the operators $O_j = \sigma^x_j$ that observes DSync. Indeed, one can verify $\tr[O_1^{\phi_1} O_2^{\phi_2} \NESSRDM_-] = \cos(\phi_1 - \phi_2 - \pi)$, suggesting a statistical propensity for the two azimuthal angles to anti-align (i.e., differ by a phase difference $\pi$). Therefore, under generic perturbations that break the strong symmetry $P$, the system must evolve to a static state $\NESSpertRDM$ that exhibits SSync. We emphasize that, specifically in this case, the SSync indicates anti-aligned azimuthal spin angles, in a picture perfectly agreeing with the DSync physics.

In the following, we characterize SSync with the commonly adopted  joint distributions of azimuthal phase angles (following the general, explicit definitions given in Appendix~\mainref{appendix:A}). They are computed for the single steady state $\NESSpertRDM$ after generic perturbations: Concretely, Eq.~\maineqref{equ:def_mixture_NESS} in the main text states that $\NESSpertRDM = c_+ \ket{v_+}\bra{v_+} + c_- \ket{v_-}\bra{v_-}$, with perturbation-dependent coefficients $c_\pm \geq 0$ satisfying $c_+ + c_- = 1$ for unit trace. Note that $\ket{v_+}\bra{v_+}$ does not contribute to SSync as it contains no off-diagonal entries in the computation basis. The only contribution comes from the $c_-\ket{v_-}\bra{v_-}$ term, so the joint distribution of the azimuthal angles $\phi_1$, $\phi_2$ reads
\eq{
    S\left(\phi_1, \phi_2\right) = \frac{1}{4\pi^2}- \frac{c_-}{64} \cos\left(\phi_1-\phi_2\right). \label{equ:spin1/2:S}
}
It follows from Eq.~\maineqref{equ:phase_diff_distribution_def} in Appendix~\mainref{appendix:A} that $S_\mathrm{d}(\phi_1') = (1/2\pi) -(\pi c_-/32)\cos\left(\phi_1'\right)$ in terms of the angular difference $\phi_1' = \phi_1 - \phi_2$.  We then find that the distribution $S_\mathrm{d}$ of the angular difference achieves maximum when $\phi_1'=\pi$: The azimuthal angles $\phi_1$, $\phi_2$ tend to differ by the phase difference $\pi$. This again agrees with the dynamical notion, where the expectation values of observables $O = \sigma^x$ at the two spins show persistent oscillations with a phase difference $\pi$.

We verify numerically that generic weak perturbations indeed lead to SSync at the angular difference $\pi$. In particular, we sample unit perturbations $\mathcal{L}_{(1)}$ from an ensemble of random Liouvillians defined in Ref.~\cite{Denisov19} and study the dynamics given by $\mathcal{L} + \eta \mathcal{L}_{(1)}$ at varying strength $\eta$. We have detailed our sampling method in Sec.~\ref{sec:numerics:sampling}. For every realization of randomness, we have found as expected a single steady state, computed its maxima calibrated by the uniform distribution, $S_\mathrm{max} = \max S_\mathrm{d} - (1/2\pi)$ [cf. Eq.~\maineqref{equ:phase_measure_def} in Appendix~\mainref{appendix:A}], and characterized the angular configuration that achieves the maxima by the function
\eq{
    \chi_2(\phi_1') = \frac{1}{\pi} \mathrm{argdist}\left(\phi_1', 0\right).
}
Here and also in the main text, we use $\mathrm{argdist}\left(\theta, \psi\right) = \min_{n\in\mathbb{Z}} \left|\theta - \psi + 2\pi n\right|$ to define a periodic angular distance on circle. Note that, in fashion similar to the $\chi$ function characterizing the angular differences of three angles [cf. Eq.~\maineqref{equ:chi_phase_diff} of the main text], here $\chi_2$ is uniformly distributed in $[0, 1]$ if the angles $\phi_1$, $\phi_2$ are uniformly i.i.d.

\begin{figure}
    \centering
    \includegraphics[width=0.75\linewidth]{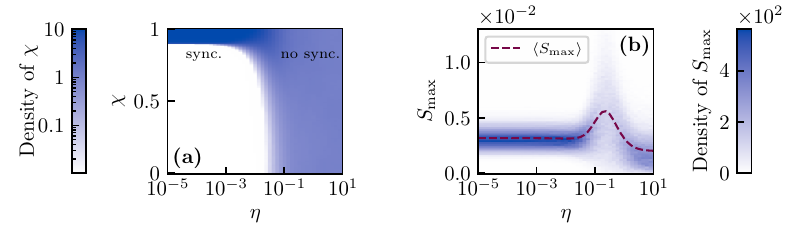}
    \caption{The model of two spin-$1/2$s~[described in Eqs.~\eqref{equ:spin1/2:H} and \eqref{equ:spin1/2:L}, with $J=1$, $\Delta = 1$, $B = 0.3$, $\gamma = 0.5$] under generic random perturbations at strenghs $\eta$. Data represent average over $10^4$ realizations of randomness. {\bf (a)} The distribution of $\chi_2(\phi_1')$ characterizing the angular difference $\phi_1'=\phi_1-\phi_2$, sampled from angular configurations such that $S_\mathrm{d}$ achieves higher than $0.95 S_\mathrm{max}$. {\bf (b)} The distribution of $S_\mathrm{max} = \max S_\mathrm{d} - (1/2\pi)$, the maximum of $S_\mathrm{d}$ calibrated by the maximally mixed state, and its average, $\braket{S_\mathrm{max}}$.}\label{fig:Smax}
\end{figure}

In Fig.~\ref{fig:Smax}, we plot the distributions of $\chi_2$ and $S_\mathrm{max}$ for increasing $\eta$. To avoid potential numerical artifacts of locating more than one local maxima of similar values, we sample $\chi_2(\phi_1')$ from those $\phi_1'$ where $S_{\mathrm{d}}(\phi_1')$ achieves above a threshold $r S_\mathrm{max}$, with $r = 0.95$. We find typically nonzero $S_\mathrm{max}$ at all the $\eta$ we have accessed, including $\eta$ that are significantly smaller than the energy scale $J$ in the unperturbed model. This suggests steady-state synchronization in a finite vicinity of the fine-tuned $\eta = 0$ point of dynamical synchronization. In addition, while we observe uniform distribution of $\chi$ indicating uniformly i.i.d. $\phi_1, \phi_2$ and thus no synchronization for $\eta/J \gtrsim 10^{-1}$, at small $\eta/J \lesssim 10^{-2}$ we find $\chi_2$ concentrated around $\chi_2 = 1$. This suggests that $\phi_1$, $\phi_2$ most likely differ by $\pi$ in the steady-state synchronization regime, as is expected from our analytical results above.

\section{Steady-state synchronization in a weakly coupled spin model}
\label{appendix:Bruder}
We give a concrete example where two decoupled spins hosting oscillating coherences feature steady-state synchronization after perturbative couplings. We adopt the example from an established model of steady-state synchronization~\cite{Bruder18_mutual} and show that the results there in some regimes can be interpreted with the mechanism for SSync that is presented in our main text.

The original model consists of two spin-1s. To fit the model under our analysis, we view the model as governed by the decoupled Hamiltonian and jump operators, on the two spins $j=\mathrm{A}, \mathrm{B}$,
\al{
    H_j &= \omega S_j^z \quad \forall j \nonumber \\
    L_\mathrm{A}^g &= \sqrt{\frac{\gamma^g_\mathrm{A}}{2}} S_\mathrm{A}^+ S_\mathrm{A}^z, \quad L_2^d = \sqrt{\frac{\gamma^d_\mathrm{B}}{2}} S_\mathrm{B}^-S_\mathrm{B}^z,
}
that give rise to the Liouvillians $\mathcal{L}_\mathrm{A}$, $\mathcal{L}_\mathrm{B}$, respectively, according to Eq.~\maineqref{equ:def_Liouvillian}. Additionally, the Liouvillian $\mathcal{L} = \mathcal{L}_\mathrm{A} + \mathcal{L}_\mathrm{B} + \mathcal{L}_{(1)}$ of the entire system contains perturbation $\mathcal{L}_{(1)}$ contributed by
\al{
    H_{(1)} &= \frac{\j \epsilon}{2}\left(S_\mathrm{A}^+ S_\mathrm{B}^- - S_\mathrm{A}^- S_\mathrm{B}^+\right) + \sum_{j = \mathrm{A}, \mathrm{B}} \Delta_j S_j^z \nonumber \\
    L_\mathrm{A}^d &= \sqrt{\frac{\gamma^d_\mathrm{A}}{2}} S_\mathrm{A}^- S_\mathrm{A}^z, \quad L_\mathrm{B}^g = \sqrt{\frac{\gamma^g_\mathrm{B}}{2}} S_\mathrm{B}^+ S_\mathrm{B}^z. \label{equ:Bruder_perturbations}
}
We focus on the weak perturbation regime defined by $\epsilon, \Delta_\mathrm{A}, \Delta_\mathrm{B}, \gamma_\mathrm{A}^d, \gamma_\mathrm{B}^g \ll \omega, \gamma_\mathrm{A}^g, \gamma_\mathrm{B}^d$, where steady-state synchronization has been observed~\cite{Bruder18_mutual}. We note however that Ref.~\cite{Bruder18_mutual} has also discussed the regime of finite perturbations, especially one where $\gamma_\mathrm{A}^d = \gamma_\mathrm{A}^g$. This regime is out of the scope of our analysis, which we have developed for weakly coupled oscillating coherences.

Our discussion focuses on the limit of weak perturbations $\epsilon, \Delta_\mathrm{A}, \Delta_\mathrm{B}, \gamma_\mathrm{A}^d, \gamma_\mathrm{B}^g \rightarrow 0$. The system in this limit possesses decoupled oscillating coherences
\eq{
    \OSC_\mathrm{A} = \ket{0}_\mathrm{A}\bra{1}_\mathrm{A}, \quad \OSC_\mathrm{B} = \ket{-1}_\mathrm{B}\bra{0}_\mathrm{B},
}
and their Hermitian conjugates. Here, in accordance with the notation in the main text, $\OSC_j$ denotes the oscillating coherence on the standalone spin $j$. Moreover, $\mathcal{L}_\mathrm{A}[\OSC_\mathrm{A}] = \j \omega \OSC_\mathrm{A}$, $\mathcal{L}_\mathrm{B}[\OSC_\mathrm{B}] = \j \omega \OSC_\mathrm{B}$, so they oscillate at the same frequency. 

Under this setup, the unperturbed system features (among others) the following zero mode,
\eq{
    \rho^{\lambda=0}_\mathrm{new} = \OSC_\mathrm{A} \otimes \left(\OSC_\mathrm{B}\right)^\dagger =  \ket{0,0}\bra{1,-1}, \label{equ:Bruder_cross_terms}
}
together with its Hermitian conjugate. These two zero modes $\rho^{\lambda=0}_\mathrm{new}$, $(\rho^{\lambda=0}_\mathrm{new})^\dagger$ are the cross terms that we have anticipated in the main text [above Eq.~\maineqref{equ:mixture_twosites}]. As is argued analytically and verified numerically in Ref.~\cite{Bruder18_mutual}, these off-diagonal entries in the computation basis indeed contribute to the steady-state measure $S_\mathrm{max}$ of synchronization [defined in Eq.~\maineqref{equ:phase_measure_def} of Appendix~\mainref{appendix:A}] in the presence of small yet nonzero $\epsilon, \Delta_\mathrm{A}, \Delta_\mathrm{B}, \gamma_\mathrm{A}^d, \gamma_\mathrm{B}^g$, which we view here as perturbative couplings. This result suggests that the cross terms in Eq.~\eqref{equ:Bruder_cross_terms} indeed enter the expansion of the unique steady state $\NESSpertRDM$ of the system under the specific form of perturbations in Eq.~\eqref{equ:Bruder_perturbations}, agreeing with what we have expected out of Eq.~\maineqref{equ:def_mixture_NESS} in the main text.

\end{document}